\documentclass[aps,reprint,amsmath,amssymb]{revtex4-2}

\pdfoutput=1 
\usepackage{graphicx}
\usepackage{amsmath}
\usepackage{gensymb}
\usepackage{upgreek}
\usepackage{xcolor}
\usepackage{tabularx}
\usepackage{array}
\newcolumntype{Y}{>{\centering\arraybackslash}X}

\begin{document}

\title{Anisotropy with respect to the applied magnetic field of spin qubit decoherence times}

\author{Yujun Choi}
\affiliation{Department of Physics, University of Wisconsin-Madison, Madison, WI 53706, United States}

\author{Robert Joynt}\email{rjjoynt@wisc.edu}
\affiliation{Department of Physics, University of Wisconsin-Madison, Madison, WI 53706, United States}

\date{\today} 

\begin{abstract}
Electron spin qubits are a promising platform for quantum computation. Environmental noise impedes coherent operations by limiting the qubit relaxation ($T_1$) and dephasing ($T_{\phi}$) times. There are multiple sources of such noise, which makes it important to devise experimental techniques that can detect the spatial locations of these sources and determine the type of source. In this paper, we propose that anisotropy in $T_1$ and $T_{\phi}$ with respect to the direction of the applied magnetic field can reveal much about these aspects of the noise. We investigate the anisotropy patterns of charge noise, evanescent-wave Johnson noise, and hyperfine noise in hypothetical devices. It is necessary to have a rather well-characterized sample to get the maximum benefit from this technique. The general anisotropy patterns are elucidated.  We calculate the expected anisotropy for a  particular model of a Si/SiGe quantum dot device.
\end{abstract}

\maketitle

\section{Introduction}

Spin qubits are a promising platform for quantum information processing machines \cite{Zwanenburg:2013p961}.  Recent progress includes the development of high-fidelity single-qubit operations \cite{Yoneda:2018p102}, coupling to resonators \cite{Mi:2018p599}, a programmable quantum processor \cite{Watson:2018p633}, universal quantum logic at 1.5K \cite{petit2020universal}, and capacitive coupling of spin qubits \cite{neyens2019measurements}.  The main obstacle, as in many quantum computing implementations, is the presence of noise that causes decoherence. The sources of decoherence are many: noise from the nuclear spin bath \cite{Coish:2009p2203, Beaudoin:2013p085320} evanescent-wave Johnson noise \cite{LangsjoenPRA2012,premakumar2017}, random telegraph and $1/f$-type charge noise \cite{Veldhorst:2014p981, eng2015isotopically, Yoneda:2018p102, connors2019low}, and noise from phonons \cite{khaetskii2001spin, Tahan:2002p035314, Amasha:2008p2332} are considered to be the main candidates.  

While the deleterious effects of noise are clearly evident, tracking down the exact sources has proved problematical.  On the positive side, the qubits themselves are very sensitive noise detectors.  They can be used to measure $T_1$, which is a measure of the noise strength at the qubit operating frequency $\omega_{op}$, and $T_2$, which is usually a measure of a weighted average of the noise strength at lower frequencies. In addition, echo techniques can be used to do noise spectroscopy.  However, while these techniques have been useful, new methods are needed to identify the type of noise and to pinpoint the positions of noise sources.
Some progress has been made recently in this direction \cite{connors2019low}, but the subject is still relatively undeveloped.

For most quantum computing platforms, $T_1$ and $T_2$ are determined by quite different physical mechanisms.  For spin qubits, the situation is somewhat different.  The qubits couple to the effective magnetic noise field, $\mathbf{B}^{(\rm{eff})}$, which is a vector.  Furthermore, there is a preferred axis set by the direction of the applied field $\mathbf{B}_0$.  We define the noise power tensor by
\begin{equation}
    \langle B_i^{(\rm{eff})} \,   B_j^{(\rm{eff})} \rangle_{\omega} =
    \int dt \: e^{i \omega t}
    \: \langle B_i^{(\rm{eff})}(t) \, B_j^{(\rm{eff})} (0) \rangle.
\end{equation}
Here the angle brackets denote a quantum and thermal average and the subscripts are Cartesian components. For the moment, let us assume that $\mathbf{B}_0 = B_0 \hat{z}$.  $T_1$ is determined by the transverse noise components: $1/T_1 \propto \langle B_x^{(\rm{eff})}  B_x^{(\rm{eff})} \rangle_{\omega_{op}} + \langle B_y^{(\rm{eff})}  B_y^{(\rm{eff})} \rangle_{\omega_{op}} $.  The dephasing rate $1/ T_\phi$, on the other hand, is determined by a weighted average of the longitudinal noise strength $\langle B_z^{(\rm{eff})}  B_z^{(\rm{eff})} \rangle_{\omega}$.  Since we have $1/T_2 = 1/2T_1 + 1/T_\phi$, all of the diagonal components of the noise tensor are accessible to experiment.  For most spin qubits $T_1 \gg T_2$, so we have the simpler equation $T_2=T_\phi$.  Henceforth we shall assume this to be the case and will refer only to $T_2$.

This vector character of the coherence time equations means that one can investigate that nature of noise in spin qubit systems by measuring the anisotropy in $T_1$ and $T_2$  as a function of the direction of the applied field.  Simply  put, if the applied field is in the direction $\mathsf R \hat{z}$, where $\mathsf{R}$ is a rotation operator that takes $\hat{z}$ into the direction with polar angles $(\theta, \phi) $, then  $1/T_1 (\theta,\phi) \propto \langle B_{\mathsf{R}\hat{x}}^{(\rm{eff})}  B_{\mathsf{R}\hat{x}}^{(\rm{eff})} \rangle_{\omega_{op}} + \langle B_{\mathsf{R}\hat{y}}^{(\rm{eff})}  B_{\mathsf{R}\hat{y}}^{(\rm{eff})} \rangle_{\omega_{op}} $ while $T_{\phi}(\theta,\phi)$ depends on $\langle B_{\mathsf{R}\hat{z}}^{(\rm{eff})}  B_{\mathsf{R}\hat{z}}^{(\rm{eff})} \rangle_{\omega}$.  The pattern in $(\theta,\phi)$ gives information about the nature of the noise sources and their positions relative to the qubit.

For charge qubits, the analog of the direction of the external field is the direction of the line in space that connects the two quantum dots. It is normally not possible to adjust this over a wide range and even in a narrow range it is unlikely to be possible in a controllable way.  As a result, the experiments we propose are only possible for spin qubits.  In this paper, we limit the analysis to single-spin devices. The idea of using anisotropy to investigate noise should also be applicable to multi-dot qubit devices such as hybrid qubits \cite{Koh:2012p250503} or singlet-triplet qubits \cite{Johnson:2005p483}. The analysis is more complicated for these cases.  For example, in the two-qubit experiment of Ref. \cite{seedhouse2021pauli}, the change in the ratio of g-factors as the field is rotated introduces additional complications.  We will not attempt any treatment of these more complex multi-qubit systems in this work, since we are mainly attempting to establish the basic principles involved, and for this purpose it is best to do the simpler cases first.  Nevertheless, it seems likely that the anisotropy in decoherence times would be a useful tool even in these more involved situations. 

We focus on the anisotropy of three different types of noise sources: charge noise, hyperfine noise, and evanescent-wave Johnson noise (EWJN) in silicon devices.  Charge noise is the most important at low frequencies and generally determines $T_2$.  EWJN is important at higher frequencies and low magnetic fields, and in many cases may determine $T_1$.  Hyperfine noise is also important, particularly in GaAs systems.  It is expected to be isotropic in $\mathbf{B}_0$, which from the viewpoint of this paper is a key experimental signature of this type of noise \cite{camenzind2018hyperfine}, as we shall discuss below.  Phonon relaxation mediated by spin-orbit coupling, in contrast, is highly anisotropic, as has been shown previously \cite{Tahan:2002p035314, Raith_PRL2012}.  This mechanism is important at higher magnetic fields $B_0$. Since we do not include it, the results we present here hold only for $B_0 \leq$ 3-4T \cite{xiao2010measurement}.  At these lower fields $T_1$ saturates. It's important to note that the anisotropy due to phonon effects is determined by the orientation of $\mathbf{B}_0$ relative to the crystal axes, while the anisotropies considered in this paper are relative to directions determined by the geometry of the device. Hot spots, where the valley and Zeeman levels cross, are also a strong source of decoherence \cite{Hao:2014p3860}.  Fortunately, it is relatively easy to avoid this by tuning of the strength of the applied field, and this would be necessary for the proposed experiment to work. In metal-oxide-semiconductor (MOS) structures the Dresselhaus interaction can be strong and anisotropic.  This gives rise to an anisotropic $T_2$ from charge noise that was measured in Ref.~\cite{ferdous2018interface}.  Here we focus on dots in heterostructures, where the angular variation of the spin-orbit coupling is expected to be much weaker. 

In most experiments on spin qubits, the relative orientation of the sample and the applied magnetic field is not allowed to vary. However, rotatable sample holders can give some variation in the angle between the growth direction and the applied field.  See, for example, Ref. \cite{sigillito2017all}.  Full coverage of the whole solid angle can be obtained from vector magnet arrangements with appropriate parameters of the magnets. Indeed, experiments to optimize qubit operation by changing the direction of the applied field have been carried out \cite{zhang2021controlling}.  The present paper can be viewed as an aid to these kinds of efforts, since the direction of maximum decoherence times can be inferred from the calculations.  Other phenomena that have been investigated by rotating the field are the variations in the Rabi frequency of multi-hole qubits in Si \cite{crippa2018electrical} and the profile of the spin-orbit interaction of a silicon double quantum dot in MOS structures \cite{tanttu2019controlling}, \cite{Marx:2020}.  

This paper will focus on experiments that use a micromagnet to provide a field gradient at the position of the dot.  The direction of the magnetization of the micromagnet can be affected by the rotation of the applied field in a way that is not well understood and that is difficult to measure.  This means that to carry out the type of experiment that is proposed here, hard magnets must be used.    

\vspace{0.2 in.}

\section{Results}
\label{sec:results}

\subsection{Relaxation time}
\label{subsec:relaxation}

The relaxation rate of a spin qubit in the noise magnetic field depends on the noise correlation function $\langle B_i^{\rm{(eff)}} (t)  B_j^{\rm{(eff)}} (0) \rangle_{\omega_{op}}$ where $\mathbf{B}^{(\rm{eff})}$ is the effective noise magnetic field, $\omega_{op}$ is the operating frequency of the qubit.  The effective noise magnetic field is any time-dependent field that couples to the spin in the usual way.  Hence this could be a physical magnetic field, a field that comes from the motion of the qubit in an inhomogeneous field, a field that results from phonons mediated by spin-orbit coupling, \textit{etc.}

$T_1$, the relaxation time, depends only on the transverse components of the correlation function.  If we define $T_1^{(i)}$ as the relaxation time when the applied field is in the $i$-direction, then:
\begin{equation}
\begin{aligned}
\frac{1}{T_1^{(x)}} &= (\frac{\mu_B}{\hbar})^2 \left[ \langle B_y^{\rm{(eff)}} B_y^{\rm{(eff)}} \rangle_{\omega_{op}} + \langle B_z^{\rm{(eff)}}  B_z^{\rm{(eff)}} \rangle_{\omega_{op}} \right], \\
\frac{1}{T_1^{(y)}} &= (\frac{\mu_B}{\hbar})^2 \left[ \langle B_z^{\rm{(eff)}} B_z^{\rm{(eff)}} \rangle_{\omega_{op}} + \langle B_x^{\rm{(eff)}}  B_x^{\rm{(eff)}} \rangle_{\omega_{op}} \right], \\
\frac{1}{T_1^{(z)}} &= (\frac{\mu_B}{\hbar})^2 \left[ \langle B_x^{\rm{(eff)}} B_x^{\rm{(eff)}} \rangle_{\omega_{op}} + \langle B_y^{\rm{(eff)}}  B_y^{\rm{(eff)}} \rangle_{\omega_{op}} \right]. \\
\end{aligned}
\end{equation}
$\mu_B$ is the Bohr magneton.  We take $g=2$.

If the applied field is in an arbitrary direction $\hat{n} = \sin\theta \cos\phi \, \hat{x} + \sin\theta \sin\phi \, \hat{y} + \cos\theta \, \hat{z}$, where $\theta$ is the polar angle and $\phi$ is the azimuthal angle, then the relaxation rate becomes
\begin{equation}
\label{eq:T1}
\frac{1}{T_1 (\theta, \phi)} = (\frac{\mu_B}{\hbar})^2  
\sum_{ij} Q^{(1)}_{ij} \, \langle B_i^{\rm{(eff)}} B_j^{\rm{(eff)}} \rangle_{\omega_{op}}
\end{equation}
where
\begin{widetext}
\[
\mathsf{Q}^{(1)}=
  \begin{bmatrix}
    \cos^2\phi \cos^2\theta + \sin^2\phi & -\cos\phi \sin\phi \sin^2\theta & -\cos\phi \cos\theta \sin\theta  \\
    -\cos\phi \sin\phi \sin^2\theta & \sin^2\phi \cos^2\theta + \cos^2\phi & -\sin\phi \cos\theta \sin\theta \\
    -\cos\phi \cos\theta \sin\theta & -\sin\phi \cos\theta \sin\theta & \sin^2\theta
  \end{bmatrix}
\]
\end{widetext}
with $\{x,y,z\}$ as the basis for the matrix $Q^{(1)}_{ij}$.
Note that if there are any nonzero off-diagonal correlation functions $\langle B_i^{\rm{(eff)}} B_j^{\rm{(eff)}} \rangle_{\omega}$ ($i\neq j$), they are also needed in the expression for the relaxation time.   

\subsection{Dephasing time}
\label{subsec:dephasing}

The calculation of the dephasing time is more complicated than that for the relaxation time, since it depends more sensitively on the full frequency spectrum of the noise and higher-level correlation functions.  For the purposes of this paper only ratios of $T_2$ for different applied field angles are important.  Hence the specific approximation used to compute $T_2$ is not so crucial.  It will be sufficient to assume that the field fluctuation obey Gaussian statistics.  Then if the applied field is in the $z$-direction, the off-diagonal components of the density matrix of the qubit decay according to the expression $\exp[-\Gamma(t)]$ with
\begin{equation}
\label{eq:T2}
\Gamma (t) = \frac{t^2}{2} (\frac{2\mu_B}{\hbar})^2 \int_{-\infty}^{\infty} d\omega \, \langle B_z^{\rm{(eff)}} B_z^{\rm{(eff)}} \rangle_{\omega} \, \mathrm{sinc}^2 (\omega t /2).
\end{equation}
The dephasing time $T_\phi$ of the qubit is obtained by solving the transcendental equation $\Gamma (T_\phi) = 1$.  Here the sinc function is defined by $\textrm{sinc} (x) = \sin x / x$.  If the applied field is in the $(\theta,\phi)$ direction, then  
\begin{equation}
\begin{aligned}
\Gamma (t) = &\frac{t^2}{2} (\frac{2\mu_B}{\hbar})^2 \,
\sum_{ij} Q^{(2)}_{ij} \\
& \times \int_{-\infty}^{\infty} d\omega \, \langle B_i^{\rm{(eff)}} B_j^{\rm{(eff)}} \rangle_{\omega} \, \mathrm{sinc}^2 (\omega t /2)
\end{aligned}
\end{equation}
where
\[
\mathsf{Q}^{(2)}=
  \begin{bmatrix}
    \cos^2\phi \sin^2\theta & \cos\phi \sin\phi \sin^2\theta & \cos\phi \cos\theta \sin\theta  \\
    \cos\phi \sin\phi \sin^2\theta & \sin^2\phi \sin^2\theta & \sin\phi \cos\theta \sin\theta \\
    \cos\phi \cos\theta \sin\theta & \sin\phi \cos\theta \sin\theta & \cos^2\theta
  \end{bmatrix}
\]
in the same basis as used for $\mathsf{Q}^{(1)}$.

The results of this subsection and the previous one make it clear that the entire tensor structure of the noise correlation function is in principle accessible simply by measuring $T_1$ and $T_2$. 

\subsection{Effective Magnetic Field Noise from a Micromagnet}
\label{subsec:micromagnet}

Single-qubit logic gates in spin systems are often implemented using a micromagnet to set up  a magnetic field gradient.  This has the unwanted complication that electric field noise moves the spin up and down the gradient, causing a time-dependent magnetic field $\mathbf{B}^{(E)}$ that can decohere the qubit.  Here we outline how this plays into the anisotropy effect.

We will take a simple model of a quantum dot in a harmonic potential.  The Hamiltonian is 

\begin{equation}
 H = -\sum_i \frac{\hbar^2}{2m_i}\frac{\partial}{\partial x_i^2} 
 + \sum_i\frac{1}{2} k_i x_i^2 - q \sum_i x_i E_i(t)
\end{equation}
where $i$ is a Cartesian index, $m_i$ and $k_i$ are the effective mass and spring constant of the electron in the $i$-direction, $q$ is the electric charge of the electron and $E_i(t)$ is the noise electric field component in the $i$-direction.  The frequency of the electric noise is much smaller than the natural frequencies of the harmonic motion, so the Born-Oppenheimer approximation applies and the effect of $E_i(t)$ is to shift the position of the minimum of the potential by $\Delta x_i(t) = q E_i(t) /2 k_i$.  The confinement in the $z$-direction is much stronger, though the effects of excited states in this potential are measurable  \cite{Scarlino:2015p106802,takeda2018optimized}.  To the level of approximation needed in this paper, we can drop the $z$ term in Eq.~\ref{eq:beff} in the potential and treat the dot as two-dimensional. 

The micromagnet sets up a static field $\mathbf{B}^m$ that varies strongly in space. The associated effective noise field that acts on the spin is
\begin{equation}
\label{eq:beff}
    B_i^{(E)}(t) = \frac{\partial B^m_i}{\partial x_j} \Delta x_j(t)
    = q \sum_j \frac{1}{2k_j}\frac{\partial B^m_i}{\partial x_j} E_j(t).
\end{equation}
The field gradient and the spring constant are device parameters.  This equation already shows that considerable device modeling is necessary to extract any interesting information about the noise field $E_i(t)$.

\subsection{Magnet Hardness}
\label{subsec:magnet}
In qubit experiments carried out to date, the micromagnet is made of pure cobalt 
\cite{tokura2006coherent, Kawakami:2014p666}. Co is hexagonal with the easy axis for magnetism along the $z$-direction.  The anisotropy parameter (the difference in energy between the $z$-axis and the $x$-$y$ plane) is about $10^6\,$J/m$^3$, which corresponds to an anisotropy field $H_A$ of about 0.5 T  when the magnetization is saturated, as it surely is at the low temperatures of the experiments.  This and shape anisotropy will mainly determine the coercive field, though sample-dependent domain wall pinning will also contribute. The anisotropy parameter quoted above is consistent with the results of Ref.~\cite{neumann2015}, though it is sometimes assumed that the magnet will rotate freely \cite{chesi2014single, ferdous2018valley}.  In any case, at the fields of present-day experiments (a fraction of 1 T), the magnet cannot be considered to be hard and some rotation of the direction of the magnetization of the Co micromagnet is certainly to be expected.  This will change the field gradient tensor that is used to calculate $T_1$ and $T_2$.  

There are three possible solutions to this problem, the first two of which involve changing the magnet.  

The first is to use a softer ferromagnet with a cubic crystal structure such as Fe, where one could expect that the magnetization follows the applied field.  This would erase much though not all of the anisotropy in the decoherence times.  It is certainly not ideal but might still give useful information about the noise sources.  

The second option is to use a harder ferromagnet, so that the field gradient tensor is fixed once and for all when the magnet is cooled in a field.  There are many possibilities, but one that suggests itself is SmCo$_5$, very close chemically to the Co magnets in current use. The anisotropy field can be as high as $H_A =$55T, an order of magnitude higher than pure Co \cite{de1998chemical}.  The magnetization would not be significantly affected by the rotation of the external field. Then the anisotropy in $T_1$ and $T_2$ is more pronounced and more information can be extracted from it.

The third possibility is to stick with the Co magnet but to use an external field that is considerably smaller ( $< 0.05$ T, say) so that the field from the micromagnet is fixed. This has the problem that the energy level splitting of the qubit becomes smaller than $k_BT$, and initializing the spin for measurements is problematic. It is possible only at temperatures of the order of 1mK, considerably lower than the temperatures in use today for these experiments.  One might still be able to do spin blockade-based measurements, however.

In this paper, we shall assume that the second alternative is chosen, since this gives the richest phenomenology, and seems feasible with fairly modest changes in fabrication techniques.  So we take the field gradients to be fixed.  Of course this means that the experiment cannot be carried out with the sample of Ref.~\cite{Kawakami:2014p666}, which uses a Co magnet and electron temperatures of order 150 mK.

\subsection{Evanescent-wave Johnson Noise}
\label{sec:ewjn}
Evanescent-wave Johnson noise (EWJN) is due to the random motion of charges in the metallic elements of the device.  This motion produces random electric and magnetic fields on the qubits in the vicinity of the metal.  For the discussion of this effect, let us take the growth direction for the device to be the $z$-direction, the distance of the qubit from the gate layer as $d$, the gate thickness as $w$, and the dielectric constant of the intervening insulating material as $\epsilon_d$.  

For the case of noise from a conducting half-space, rather simple formulas are available \cite{henkel1999, LangsjoenPRA2012}.  In most Si/SiGe heterostructure and Si MOS devices the gates form sheets of metal that can be approximated as a uniform layer from the standpoint of noise production.  Thus the theory of EWJN from a film with a finite $w$ is more appropriate.  It has been worked out in detail \cite{langsjoen2014}, though the results are somewhat complicated, and depend on whether we consider electric field noise or magnetic field noise. For the values of $d$ of interest to us, electric field noise is slightly enhanced for the film case as compared to the half-space case, while the opposite is true for the magnetic field noise, and the effect of finite $w$ is larger.  A very good approximation is to use the half space formula for the electric noise and a modified formula for the magnetic case. 

Given these considerations, the noise correlation functions for the electric field are
\begin{equation}
    \langle E_z  E_z \rangle_{\omega_{op}} = \frac{\hbar \omega_{op} \epsilon_d \epsilon_0}{2 \sigma d^3}
    \coth\frac{\hbar \omega_{op}}{2 k_B T}.
\end{equation}
Here $\sigma$ is the conductivity.  The other elements of the noise tensor are $\langle E_x  E_x \rangle_{\omega_{op}} = \langle E_y  E_y \rangle_{\omega_{op}} = (1/2) \langle E_z E_z \rangle_{\omega_{op}}$, while the off-diagonal elements of the tensor vanish. This electric noise is converted into effective magnetic field noise using the techniques of the previous section.  

The magnetic EWJN correlation function is given by 

\begin{equation}
\label{eq:mag}
    \langle B_z  B_z \rangle_{\omega_{op}} = \frac{\hbar \omega_{op} \mu_0 \sigma w}{8 d^2}
    \coth\frac{\hbar \omega_{op}}{2 k_B T}.
\end{equation}
This is reduced from the half-space result by a factor of $w/d$. The other elements of this noise tensor are $\langle B_x  B_x \rangle_{\omega_{op}} = \langle B_y  B_y \rangle_{\omega_{op}} = (1/2) \langle B_z B_z \rangle_{\omega_{op}}$, while the off-diagonal elements of the tensor vanish. Unlike the electric noise, magnetic EWJN acts directly on the qubit spin to produce decoherence.

\subsection{Charge Noise Sources}

The exact nature of the low-frequency charge noise remains controversial.  There are two leading models for the source of the two-level systems (TLS) that give rise to this noise.

The first model is the TLS proposal of Anderson, Halperin and Varma, and independently Phillips, of 1972 \cite {anderson1972anomalous, phillips1972tunneling}.  The picture, shown in Fig.~1, is that the noise source is the motion of some atom or group of atoms in a potential that supports bistability in some range of parameters.  The motion also produces a fluctuating dipole moment, which we take as $\pm \mathbf{p}$ in the two stable positions.  For the most part, this idea has formed the conceptual background of the field in physics experiments for the last half century.  We will assume that the orientation of these dipoles is uniformly random on the unit sphere, and that they are distributed uniformly in space in the oxide layer above the qubits. We call this the random dipole model. 

The second model is the related but physically quite distinct idea of McWhorter \cite{mcwhorter1957semiconductor}, in which a conducting layer serves as a reservoir for electron traps near the surface of the layer whose energies are close to the Fermi level of the layer.  Electrons from the reservoir can hop on and off, again changing the distribution of charge in the system, the change being well approximated by a fluctuating dipole perpendicular to the layer.  In this case the fluctuation is between a zero and a fixed nonzero value of the dipole moment, while in the dipole model the fluctuation is between two different nonzero values.  The trap-type TLS is in fact widely thought to be the most important for the noise in field-effect transistors in the engineering community.  However, as in the case of the dipole model, real proof of the details of the model is hard to come by. We call this the trap model.  An illustration of the model is given in  Fig.~1(b).

It is evident that the two models are not easy to distinguish experimentally, since they will both give random telegraph noise with a distribution of switching rates, and reasonable assumptions about the distribution will lead to $1/f$ type noise. They differ in the orientation of the effective dipoles, however, suggesting that an experiment that can detect anisotropy in the noise correlation tensor will distinguish the two models.  This forms a chief motivation for the current work.

 \begin{figure}[t]
     \centering
     \includegraphics[width=0.48\textwidth]{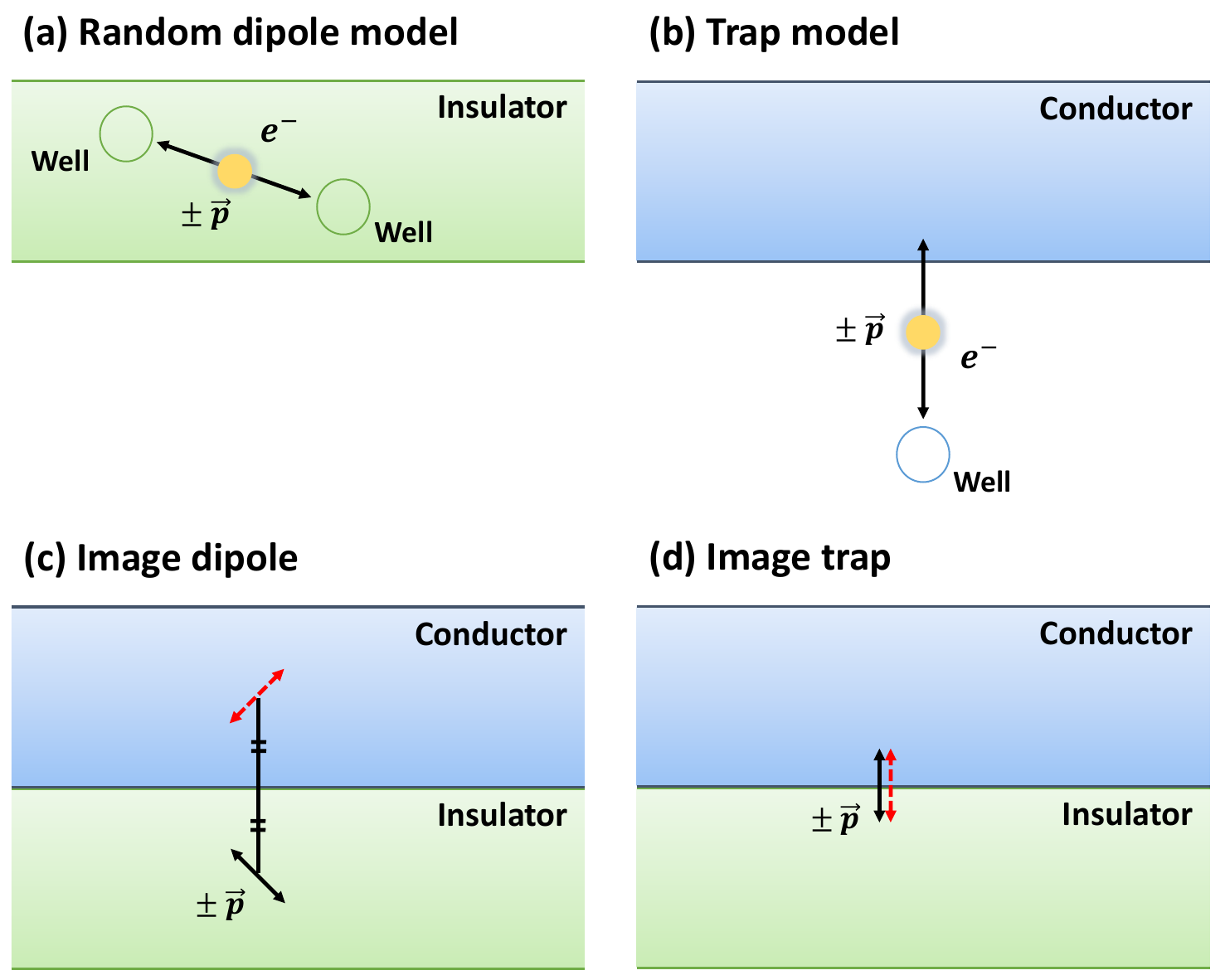}
     \caption{Conceptual diagram of random dipole model and trap model. (a) and (b): A charge moves back and forth between bistable potential wells in the random dipole model while between the Fermi surface of a conductor and a trap of an insulator in the trap model.  Screening effect of metallic gates can be taken into account by adding image charges in the conductor.  (c) and (d): Resultant image dipole and image trap are depicted as red dashed arrows respectively.}
     \label{fig:models}
 \end{figure}

\subsection{Source Positions}

Anisotropy in the decoherence rates can come from several sources. We have seen that the noise tensor from EWJN can itself be anisotropic, and the dipole and trap models also have characteristic anisotropy signatures.  In addition to this, it is possible for charge noise sources to be clumped, either from a tendency to adhere to different device elements, or, particularly in the case of only a few sources, to cluster by random chance.  Of course if the noise is coming from a certain direction this is also a source of anisotropy.

This leads to distinguishing four models all together, which we call the uniformly distributed dipole model (UD), uniformly distributed trap model (UT), localized cluster dipole model (CD) and localized cluster trap model (CT).  The U-type models assume that the sources are many in number and uniformly distributed, while the C-type models assume that the sources are relatively few in number.  The total number of sources is of course also very important to determine \cite{connors2019low}.

The U- and C- type models represent limiting cases of very many and just a few closely spaced noise sources, respectively.  Of course it is not possible to rule out in advance a lumpy set of say 10 to 100 noise sources.  The current computational method would need to be developed considerably further to become a useful characterization tool in this difficult intermediate case.  In particular, multiple qubits and cross-correlation functions among them would most likely be needed.

\subsection{Noise Correlation Functions} 
The noise correlation functions for all models are calculated as follows.

First we note that because of the metallic elements in the device, it is important to include screening of the electric noise.  Electric dipoles near a metal surface at $z=0$ are screened in an anisotropic fashion.  A dipole oriented perpendicularly to the surface is anti-screened, since the image dipole is in the same direction as the original one.  This is in sharp contrast to a dipole oriented parallel to a metal surface, which is strongly screened, with the image dipole opposite in direction to the original one \cite{connors2019low}.  These image charge effects are taken into account in our calculations by adding an image dipole $\mathbf{p}_{im}$ to the bare dipole $\mathbf{p}_{0}$.  If the bare dipole $\mathbf{p}_0=(p_{x0},p_{y0},p_{z0})$ is located at $\mathbf{r}=(x,y,z)$ relative to a metallic layer at $z=0$ the image dipole is located at $(x,y,-z)$ and its moment is given by $\mathbf{p}_{im}=(-p_{x0},-p_{y0},p_{z0})$. Examples of a bare dipole and corresponding image dipole for the random dipole model and trap model are shown in Fig.~1(c) and Fig.~1(d).

Let the qubit be at the point $\mathbf{r}$, so we are interested in the correlation function $\langle E_i(\mathbf{r},t)  E_j(\mathbf{r},0) \rangle$. Let the dipole be at $\mathbf{r}'$ and the root-mean-square dipole strength be $p_0$.  Define $\mathbf{R}' = \mathbf{r} - \mathbf{r}'$.  Then 
\begin{equation}
\label{eq:eit}
 E_i(\mathbf{r},t) = \frac{1}{4\pi \epsilon_0} \sum_{\mathbf{r}'} \sum_k  \frac{3 R'_k p_k(t) R'_i - R'_k R'_k p_i(t)}{|\mathbf{R}'|^5}.
 \end{equation}
 
 The dipoles, (except for direct and image pairs) are assumed to be statistically independent, so their correlation function is
 \begin{equation}
 \label{eq:pp}
     \langle p_i(\mathbf{r}',t) p_j(\mathbf{r}',0) \rangle =
      \delta_{mod} \, p_0^2 \, g(t).
 \end{equation}
 where the model-dependent factor $\delta_{mod} = \delta_{ij}/3$ for the random dipole model, $\delta_{mod} = \delta_{iz} \delta_{jz}$ for the trap model, and $\delta_{ij}$ is the Kronecker delta.
 Here $g(t)$ is the time correlation function for a single dipole.  We substitute Eqs.~\ref{eq:eit} and \ref{eq:pp} into the definition of the electric field correlation function and, after some calculation and a time Fourier transform, find
 \begin{equation}
 \label{eq:ee1}
  \langle E_i(\mathbf{r})  E_j(\mathbf{r}) \rangle_{\omega} =
  \frac{1}{48 \pi^2 \epsilon_0^2} \rho_v p_0^2 \, g(\omega) \int d^3r'
  \frac{3 R'_i R'_j + \delta_{ij} |\mathbf{R}'|^2}{|\mathbf{R}'|^8}
 \end{equation}
 for the UD model and
  \begin{equation}
  \begin{aligned}
   \label{eq:ee2}
  \langle E_i(\mathbf{r})  E_j(\mathbf{r}) \rangle_{\omega} = \,
  &\frac{1}{16 \pi^2 \epsilon_0^2} \rho_a p_0^2 \, g(\omega) \int d^2r'  |\mathbf{R}'|^{-10} \\
  &\times \left[
  9 R_z^{\prime 2} R'_i R'_j - 3R'_z R'_j |\mathbf{R}'|^2 \delta_{iz} \right. \\
  &- \left. 3R'_z R'_i |\mathbf{R}'|^2 \delta_{jz} + |\mathbf{R}'|^4 \delta_{iz}\delta_{jz} \right]
  \end{aligned}
 \end{equation}
 for the UT model.
 Here $\rho_v$ is the volume density of dipoles, $\rho_a$ is the areal density of dipoles, and $g(\omega)$ is the Fourier transform of $g(t)$. Both direct and image pairs are included in the rate formulas \ref{eq:T1} and \ref{eq:T2}. 
 
 For charge noise $g(\omega)$ is often of the $1/f$ type.  However, one of the important advantages of the experiments described in this paper is that we can investigate the sources of noise using spatial and geometric information alone, and the frequency spectrum of the noise is less important, a point we will return to below.   
 
 For the CD and CT models, the passage to an integral as in Eqs.~\ref{eq:ee1} and \ref{eq:ee2} is not possible, and the sum in Eq.~\ref{eq:eit} must be performed explicitly.  
 
We now use the above theory to make predictions for the coherence time anisotropy for a model device. We have chosen the device parameters from the sample used by Kawakami  \textit{et al}. \cite{Kawakami:2014p666} because it has been particularly well-characterized: $T_1$ and $T_2$ were measured, the distance from the qubits to gates is accurately known as $d=137\,$nm, and, most importantly, the magnetic field gradient tensor created by the micromagnet was simulated in detail.  We take the tensor as fixed and independent of the direction of the applied field, though we repeat that this would be unlikely if a Co magnet is used.  Parameters for the device are given in the Methods section.

\subsection{Anisotropy of T$_2$}
\label{subsec:t2}

 \begin{figure*}[t]
     \centering
     \includegraphics[width=\textwidth]{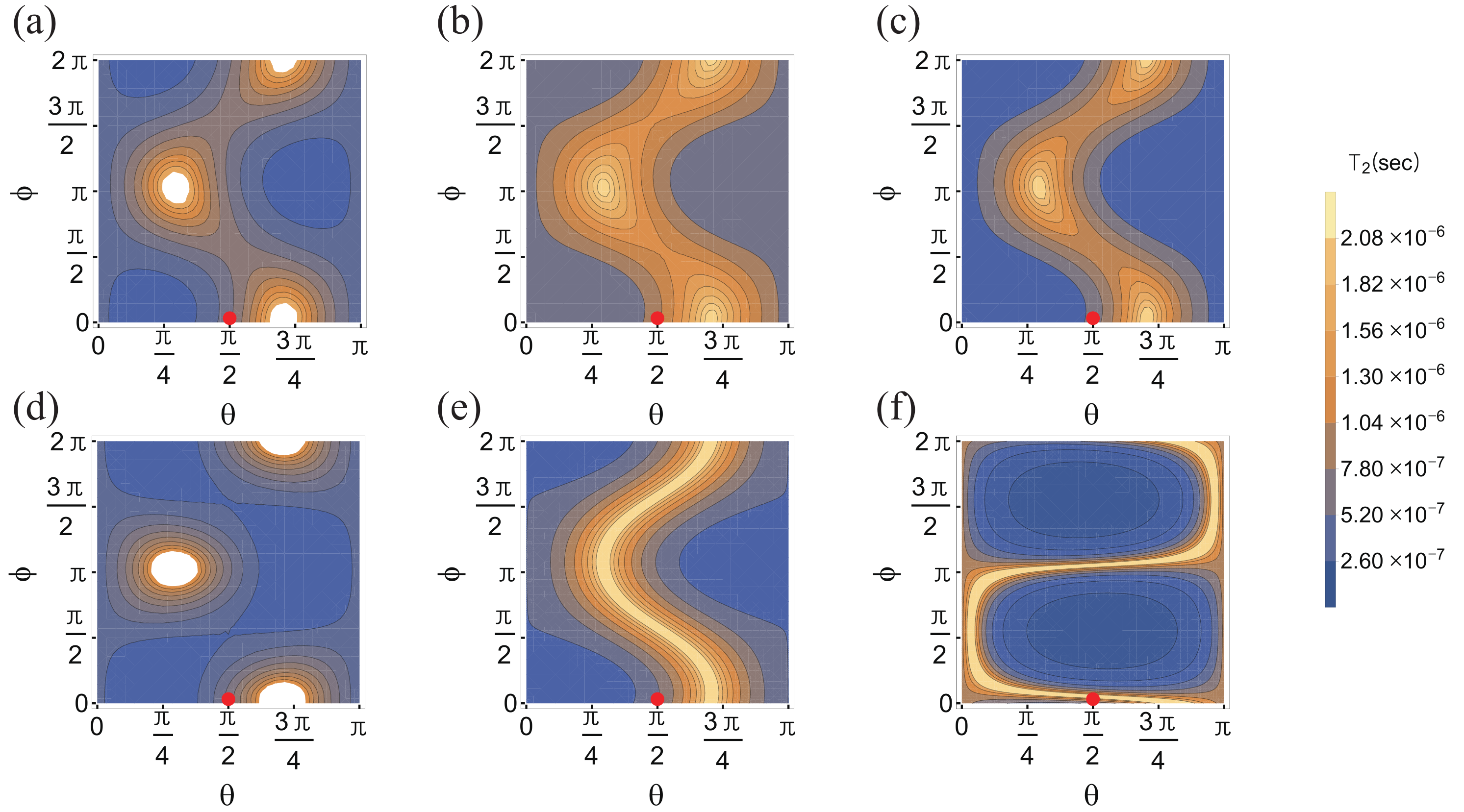}
     \caption{The anisotropy maps of decoherence time $T_2$. $x$ and $y$ axis are the polar angle $\theta$ and the azimuthal angle $\phi$ with respect to the device's $z$-direction, respectively. The models used for simulations are: (a) uniformly distributed random dipoles (UD), (b) uniformly distributed traps (UT), (c) and (d) single dipole cluster (CD) located at $(x, y, z) = (37,0,37)\,$nm and $(x, y, z) = (0,37,37)\,$nm respectively, (e) and (f) single trap cluster (CT) located at $(x, y, z) = (37,0,137)\,$nm and $(x, y, z) = (0,37,137)\,$nm respectively. The qubit is located at the origin. The applied field direction ($\theta$,$\phi$) = ($\pi/2$,$0$) used in the experiment is indicated by the red dot. For the uniform distribution models in (a) and (b), the volume density $\rho_v$ and areal density $\rho_a$ are respectively used as fitting parameters to match $T_2 (\pi/2, 0)$ to the experimental value, $840\,$ns, and for single cluster models in (c)-(f) the dipole strength $p_0$ is used as a fitting parameter.}
     \label{T2n}
 \end{figure*}

The goal of this section is to determine signatures of the different types of noise in $T_2$ data.  We first show that EWJN is not important for $T_2$.  Then we point out that hyperfine noise is isotropic and estimate its magnitude.

In the experiment of Ref.~\cite{Kawakami:2014p666}, $T_2^*$ of the device is measured to be $840\pm70\,$ns and corresponds to $T_2^{(x)}$ of our simulation because the applied magnetic field $B_0$ is in the $x$-direction. For EWJN, the dephasing rate in the $i$-direction is
\begin{equation}
\frac{1}{T^{(i)}_\phi} =   2\pi k_B T (\frac{2\mu_B}{\hbar})^2 \lim_{\omega \rightarrow 0} \left[\frac{1}{\omega} \langle B_i^{\rm{(eff)}} B_i^{\rm{(eff)}} \rangle_{\omega} \right] .
\end{equation}
This is the dephasing time for the exponential regime when the time $t \gg \hbar / k_BT.$  The calculated $T^{(x)}_2 = 1.19\,$s from EWJN is six orders of magnitude larger than the experimental one. Thus the dominant mechanism for the decoherence of the qubit should be charge noise, not EWJN.

The isotropic hyperfine noise is taken into account by estimating it from other experiments.  It should not differ too much from one device to another.  The material in the case study device is natural silicon. If the decoherence rate in isotopically purified silicon \cite{Yoneda:2018p102} is subtracted from that in natural silicon \cite{Takeda:2016p1600694}, we find a hyperfine contribution to the rate of $1/T_2^{hyper} = (1.83\upmu$s$)^{-1} - (20.4\upmu$s$)^{-1} = (2.01\,\upmu$s$)^{-1}$.  $1/T_2^{hyper}$ is then simply added to the dephasing rates from charge noise. Because of its isotropy, its effect is to smooth the resulting plots.  

For charge noise, $\Gamma(t)$ for the applied field along the $(\theta, \phi)$ direction can be written as
\begin{equation}
\label{eq:Gamma}
\Gamma (t; \, \theta, \phi) = \sum_{ij} Q^{(2)}_{ij}  \left[ \gamma_{2x} (t)  \frac{\partial B_{i} }{\partial x} \frac{\partial B_{j} }{\partial x} + \gamma_{2y} (t) \frac{\partial B_{i} }{\partial y} \frac{\partial B_{j} }{\partial y} \right]
\end{equation}
where $\gamma_{2x} (t)$ and $\gamma_{2y} (t)$ are the prefactors related to the gradients in $x$- and $y$-direction respectively.  

Now we assume that the TLS noise is a Poisson process with an exponential time correlation functions with characteristic relaxation time $\tau$ and carry out the necessary integrations.  For the UD model we have
\begin{equation}
\label{eq:ud}
\begin{aligned}
    \gamma_{2x} (t) = \gamma_{2y} (t) = \; &(\frac{2\mu_B}{\hbar})^2 (\frac{q}{2m w_{orb}^2})^2  \frac{\rho_v p_0^2}{576 \pi \epsilon_0^2} \left( \frac{1}{l^3} - \frac{1}{d^3} \right) \\
    &\times 2\pi \tau (t+(e^{-t/\tau}-1)\tau).
\end{aligned}
\end{equation}
For the UT model we find
\begin{equation}
\label{eq:ut}
\begin{aligned}
    \gamma_{2x} (t) = \; &(\frac{2\mu_B}{\hbar})^2 (\frac{q}{2m w_{orb}^2})^2 \frac{(9\pi+6) \rho_a p_0^2}{512 \pi^2 \epsilon_0^2 d^4} \\
    &\times 2\pi \tau (t+(e^{-t/\tau}-1)\tau), \\
    \gamma_{2y} (t) = \; &(\frac{2\mu_B}{\hbar})^2 (\frac{q}{2m w_{orb}^2})^2 \frac{(9\pi-6) \rho_a p_0^2}{512 \pi^2 \epsilon_0^2 d^4} \\
    &\times 2\pi \tau (t+(e^{-t/\tau}-1)\tau).
\end{aligned}
\end{equation}
The temporal part of $\gamma_{2x}$ and $\gamma_{2y}$ results from the integration of the product of a Lorentzian $g(\omega) = 2\tau/(1+(\omega \tau)^2)$, and $\mathrm{sinc}^2 (\omega t/2)$.  Those equations are obtained by converting electric field noise correlations (Eq.\ref{eq:ee1} or Eq.\ref{eq:ee2}) into effective magnetic field correlation using Eq.\ref{eq:beff} (See details in Supplementary Notes).  It is important to note first that the details of the noise spectrum and thus the choice of an exponential correlation are not crucial for the anisotropy patterns, since they depend only on ratios of noise strengths.  On the negative side, if some parameter of the noise such as $\tau$ itself depends on position in the sample, then the extraction of useful information from the analysis of the data would become far more complicated. 

$\rho_v$ and $\rho_a$ are poorly known, so we use them as fitting parameters. $T_2^* = 840 \, $ns was measured for only a single direction of the field, indicated by the red dots in Fig.~2.  This yields  $\rho_v = 2.93 \times 10^{20}\, \mathrm{m}^{-3}$ and $\rho_a = 2.66 \times 10^{11}\, \mathrm{m}^{-2}$ for Fig.~2(a) and Fig.~2(b), respectively.

The anisotropy maps of $T_2$ for the various models are shown in Fig.~2.  There is some redundancy in the maps since they are symmetric under the transformation $\theta \to \pi - \theta$ and $\phi \to \pi + \phi$, stemming from $T_2 (\mathbf{B}_0) = T_2 (-\mathbf{B}_0)$. This same redundancy also arises in the anisotropy maps of $T_1$ in Fig.~3. 
We have chosen to show the full angular ranges since in some instances the topology of the function is clearer this way.  The number of maxima ($N_{max}$), minima ($N_{min}$), and saddle points ($N_s$) in each map are shown in Table I.  When the dephasing and relaxation times in the maps are continuous functions without higher-order critical points, Morse theory can be applied. In particular for such functions on a 2D sphere (genus zero), they follow the relation: $N_{max} + N_{min} = N_s + 2$. Table I shows how the topology of the function can change when parameters affecting the noise are varied. For the maps of  2(e) and  2(f), Morse theory cannot be applied because the ridges in the maps correspond to a line of higher-order critical points.

The background $T_2^{hyper} = 2.01\,\upmu$s sets an upper bound on the plotted values in Fig.~2.  The white regions in Fig.~2(a) and Fig.~2(d) represent angular regions where the charge noise contribution is negligible and this upper bound is reached.

Fig.~2(a) shows the results for the UD model and Fig.~2(b) for the UT model. The horizontal (vertical) axis denotes polar (azimuthal) angle with respect to the device's $z$-direction. The key feature of these two models is that anisotropy of $T_2$ results only from the magnetic field gradients.  The patterns are not too dissimilar, with the ratio between maximum and minimum values being around 3 for the UD model and 2 for the UT model.  The main difference between the UD and UT models is that the peaks and valleys are broader in the UT model.  In the UD model, the dipoles are oriented randomly, while in the UT model they are in the $z$-direction.  The differences in the anisotropy maps between UD and UT can be traced back to the different behavior of electric field lines from these two different types of sources.
However, this does not manifest itself in a simple way because of the complexity of the gradient tensors that mediate the electric noise.  Because of that, it is difficult to develop much physical intuition about the distinction between UD and UT charge noise sources from inspection of the anisotropy maps, and it appears that a full calculation is necessary to test the differences between the two noise models.

The anisotropy maps for the CD model, a localized dipole cluster, are shown in Fig.~2(c) and in Fig.~2(d).  The cluster is located at $(x, y, z) = (37,0,37)\,$nm and $(x, y, z) = (0,37,37)\,$nm respectively. The maps for the CT model, a localized trap cluster, are shown in Fig.~2(e) and  Fig.~2(f). The trap is located at $(x, y, z) = (37, 0, 137)\,$nm and $(x, y, z) = (0,37,137)\,$nm respectively. In both CD and CT model, the qubit is located at the origin. Thus Fig.~2(c) is directly comparable to Fig.~2(e) and Fig.~2(d) is directly comparable to Fig.~2(f).  The overall dipole strength $p_0$ is used for the fitting parameter of these single cluster models, once more by using the experimental value measured at the red point. Figs.~2(c)-2(f) exhibit more anisotropy relative to the uniform distribution models. This is expected since the localization of the source itself introduces anisotropy.  On the other hand, one might expect that C- and U-type models would be easy to distinguish because of a simpler azimuthal dependence for the latter. But once more because of the mediation of the noise by the complicated field gradient tensors, such simple expectations are not borne out. 

The difference between the CD and CT models lies in the dipole orientation. In the CD model, it is assumed that the cluster contains dipoles of all orientations and the noise electric field is averaged over the solid angle.  This washes out the anisotropy to some extent, but the pattern still depends on the direction of the line connecting the dipole and the qubit.  The distance between the dipole and qubit just changes the overall magnitude of $T_2$. In the CT model, however, the trap generates a noise electric field with more directionality, so the overall anisotropy patterns are sharper and both the direction and the distance are important.

Comparing Fig.~2(c) to Fig.~2(d) and Fig.~2(e) to Fig.~2(f) indicates that the source position has a large effect.  To understand this in more detail, let us focus on the CT model in Fig.~2(e) and Fig.~2(f). Note that $\partial B_{x}/\partial x$, $\partial B_{z}/\partial x$, and $\partial B_{y}/\partial y$ are an order of magnitude greater than the other gradient terms. In Fig.~2(e), the electric field at the qubit has only $x$ and $z$ components. The $x$ component contributes to $1/T_2$ after multiplication by $\partial B_{x}/\partial x$ and $\partial B_{z}/\partial x$. Thus small $T_2$ is expected when the applied field is in the $x$ and $z$ directions, which can be identified on the map with $(\theta, \phi) = (\pi/2,0)$ and $(\theta, \phi) = (0,0)$, respectively. On the other hand, in Fig.~2 (f), the electric field at the qubit has only $y$ and $z$ components. The leading contribution to $1/T_2$ is the result of the product of the $y$ component and $\partial B_{y}/\partial y$. A small $T_2$ is expected with the applied field is in the $y$-direction, which is seen at the point $(\theta, \phi) = (\pi/2,\pi/2)$ on the map.  Thus for the distinction between CD and CT models, some relatively simple physical considerations can help to decipher the anisotropy map.

It is important to point out that if relatively few two-level systems contribute to the dephasing of the qubit, as is often hypothesized based on deviations for power-law spectra \cite{connors2019low,ahn2021microscopic}, the anisotropy can be used to determine the source position and to distinguish between random dipole and trap models for the charge noise.  The present method can be extended to models with very few sources by eliminating the averaging we have performed, but the analysis quickly becomes complicated.

 \begin{table}[t]
     \centering
     \includegraphics[width=0.48\textwidth]{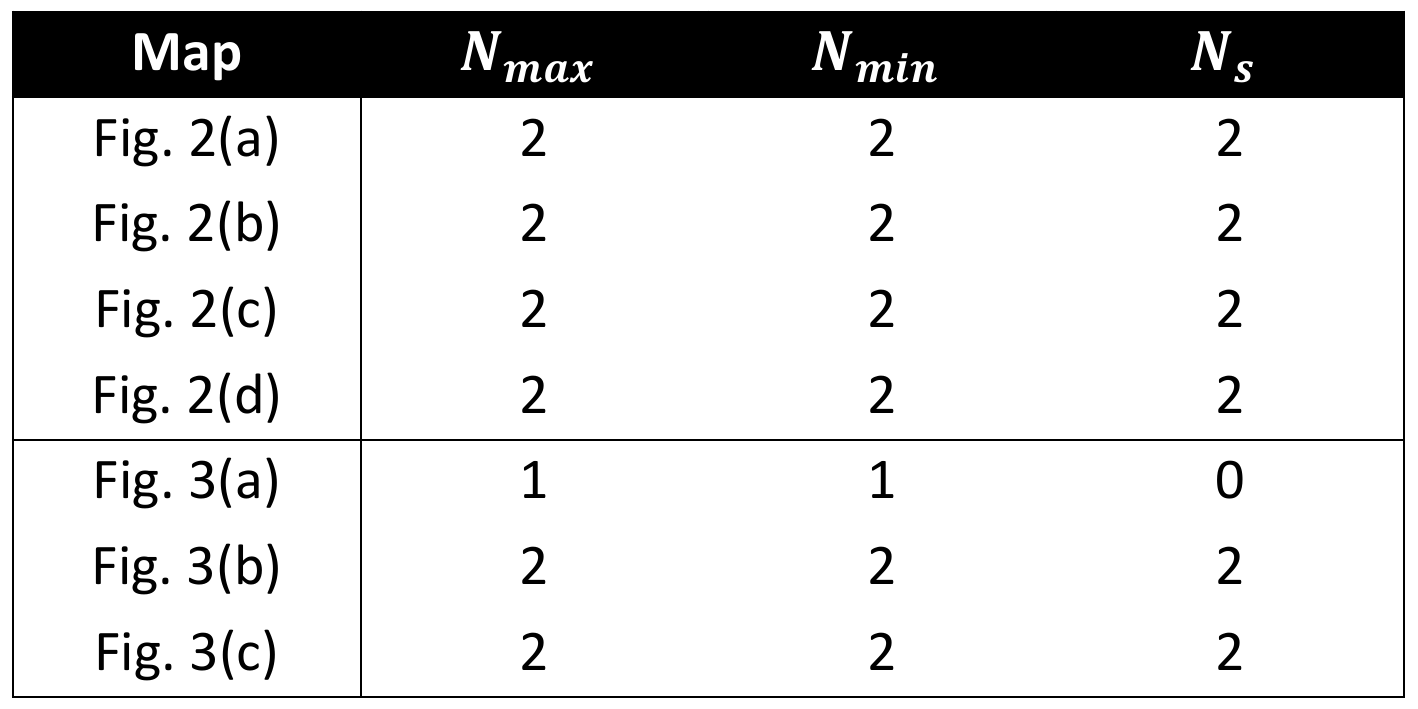}
     \caption{The number of maxima, minima, and saddle points of each anisotropy map. $N_{max}$ is the number of maxima, $N_{min}$ is the number of minima, and $N_s$ is the number of saddle points. The numbers for Figs. 2(e) and 2(f) are not included since they have higher-order critical points.}
     \label{morse}
 \end{table}

\subsection{Anisotropy of T$_1$}
\label{subsec:t1}

 \begin{figure*}[ht]
     \centering
     \includegraphics[width=\textwidth]{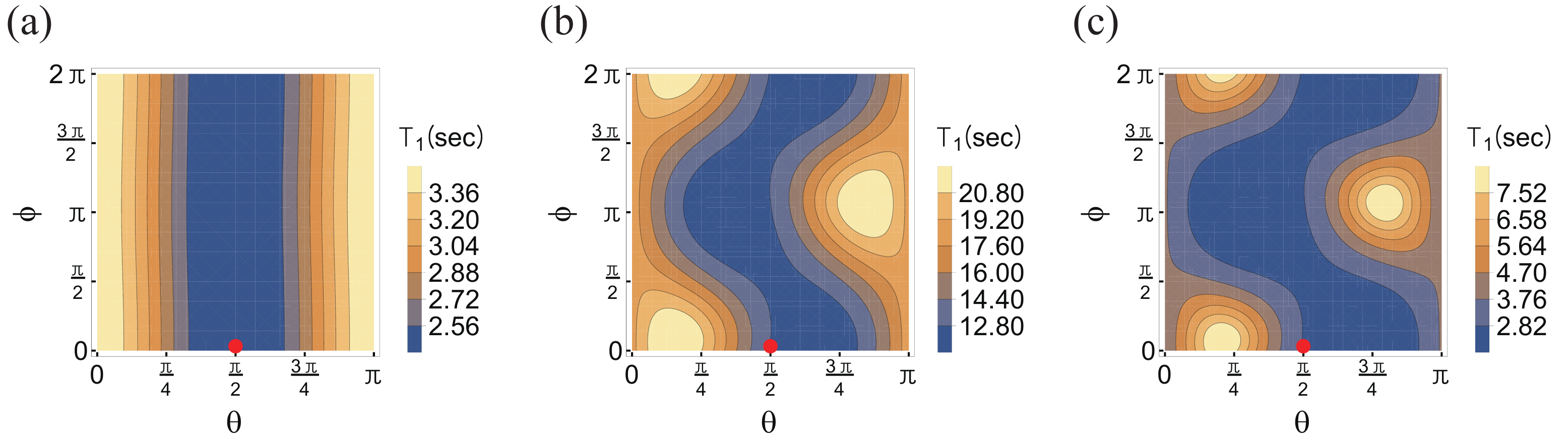}
     \caption{The anisotropy maps of relaxation time $T_1$.  $x$ and $y$ axis are the polar angle $\theta$ and the azimuthal angle $\phi$ with respect to the device's $z$-direction, respectively.  The conductivity of gates are (a) $\sigma = 2\times10^8\,$S/m, (b) $\sigma = 2\times10^7\,$S/m, and (c) $\sigma = 2\times10^6\,$S/m. The applied field direction ($\theta$,$\phi$) = ($\pi/2$,$0$) used in the experiment is indicated by red dot.}
     \label{T1n}
 \end{figure*}

In the experiment, $T_1$ of the device is in the order of $1\,$s.  To estimate the contribution of charge noise, we use the results from \ref{subsec:relaxation} together with the determination of densities from $T_2$.  This leads immediately to an estimate in the range of $10^9\,$s so we conclude that charge noise is not important for spin relaxation in the single-qubit system considered here.  To exclude phonon relaxation we need to stipulate for the moment that the external field strength is less than about $1\,$T.  This leaves EWJN as the dominant mechanism.  

The relaxation rate with applied field direction in $(\theta, \phi)$ for EWJN can be written as
\begin{equation}
\frac{1}{ T_1 (\theta, \phi) } = \alpha + \beta \sum_{ij} Q^{(1)}_{ij}  \left[  \frac{\partial B_{i} }{\partial x} \frac{\partial B_{j} }{\partial x} +  \frac{\partial B_{i} }{\partial y} \frac{\partial B_{j} }{\partial y} \right]
\end{equation}
where
\begin{equation}
\begin{aligned}
\alpha &\equiv \frac{\hbar \omega_{op} \mu_0 \sigma w}{ 4 d^2} \coth \frac{\hbar \omega_{op}}{2 k_B T},\\
\beta &\equiv (\frac{q}{2m \omega_{orb}^2})^2 \frac{\hbar \omega_{op} \epsilon_d \epsilon_0}{4 \sigma d^3} \coth \frac{\hbar \omega_{op}}{2 k_B T}.
\end{aligned}
\end{equation}

The first term represents the direct effect of magnetic field noise while the second term is due to the electric field noise that is converted to effective magnetic field noise by the field gradient created by the micromagnet.  The angular variation in $Q_{ij}^{(1)}$ and the rather anisotropic character of the magnetic field gradients imply that the anisotropy of $T_1$ will be intensified when the electric term is bigger than the direct magnetic term.  Noting that $\alpha \sim \sigma$ and $\beta \sim 1/\sigma$, we see that varying $\sigma$ will change the anisotropy pattern of $T_1$.  As noted above, $\sigma$ was not measured in the experiment and it makes sense to vary it to investigate the angular dependence of $T_1$.

The anisotropy map of $T_1$ is shown in Fig.~3 with (a) $\sigma = 2 \times 10^8\,$S/m, (b) $\sigma = 2 \times 10^7\,$S/m, and (c) $\sigma = 2 \times 10^6$\,S/m. The anisotropy pattern in  Fig.~3(a) is fairly simple because the magnetic noise is dominant and the direct magnetic EWJN itself is not very anisotropic, as can be seen from Eq.~ \ref{eq:mag} and the text following it.  The anisotropy is increased as shown in  Fig.~3(b) where the magnetic noise is somewhat more comparable to the electric noise. The anisotropy becomes even larger in  Fig.~3(c) where the magnetic noise is one order of magnitude smaller than the electric noise. This pattern looks like the reversal of the anisotropy map of $T_2$ in Fig.~2(a) and Fig.~2(b). This is natural since $T_2$ of a spin qubit is due to longitudinal noise while $T_1$ is due to transverse noise. From practical point of view, the qubit performance would be improved when the applied field direction is set to the angles that give maximal $T_2$ (in the case of $T_{\phi} \ll T_1$).

When the strength of the applied magnetic field is increased, there is a crossover from EWJN- to phonon-dominated spin relaxation.  The anisotropy maps for phonons was worked out in Ref.~\cite{Tahan:2002p035314}. They are determined by the orientation of the field relative to the crystal axes, not axes coming from the the device geometry.  Hence we expect sharp changes in the anisotropy map as the field is increased beyond about 3 T.

\section{Discussion}
\label{sec:conclusion}

The anisotropy pattern for the dephasing time of a spin qubit comes from the combination of the magnetic field gradient and the noise electric field, the latter being determined by the configuration of noise dipoles. By introducing a vector magnet in a quantum dot device, noise characteristics such as noise dipole type and/or spatial distribution of noise dipoles can be experimentally investigated. Another way to obtain similar information is to exploit a controllable magnetic field gradient for a spin qubit on a nitrogen vacancy center in a diamond \cite{zhang2017selective,zimmermann2020selective}. In this case, the gradient can be varied instead of the direction of applied magnetic field to study noise characteristics.

The anisotropy maps of relaxation times can be explained by a combination of direct magnetic noise and indirect electric noise. The magnetic noise part resulting from EWJN is isotropic in the $x-y$ plane in typical device structures.  The electric noise is mediated by magnetic field gradients, which is the only source of anisotropy. As a result, the anisotropy gets bigger as the influence of the electric noise part increases.

To summarize, we have shown anisotropy in relaxation times and dephasing times using device parameters taken from quantum dot of Kawakami \textit{et al}. Making the anisotropy map can help to understand the noise mechanisms.  Specifically, this will benefit the understanding of solid state quantum processors where the causes of noise are still being investigated. Our work contributes to science in the noisy intermediate-scale quantum era by suggesting a new experimental method for noise characterization of spin qubit devices.


\section{Methods}
\label{sec:method}
\subsection{Device parameters}
The parameters for the device of Ref.Fig.~\cite{Kawakami:2014p666} are as follows.  The field gradients at the qubit in units of mT(nm)$^{-1}$ are $\partial B_{x}/\partial x = -0.20$, $\partial B_{y}/\partial x = -0.05$, $\partial B_{z}/\partial x = -0.27$, $\partial B_{x}/\partial y = -0.03$, $\partial B_{y}/\partial y = 0.18$, and $\partial B_{z}/\partial y = -0.02$.   The $z$-direction of the device is taken to be the growth direction.  The variation in the $z$-direction is not needed in the two-dimension approximation we are using.  Other important parameters are the thickness of the aluminum oxide layer $l = 100\,$nm, the dielectric constant $\epsilon_d = 13.05$ for $\rm{Si}_{0.7}\rm{Ge}_{0.3}$ \cite{levinshtein2001properties}, and the transverse effective mass  $m = 0.19 \, m_e= 1.73 \times 10^{-31}\,$kg.  The lowest orbital excitation frequency is taken as $\omega_{orb}=6.84 \times 10^{11}\,\mathrm{s}^{-1}$ and it is related to the spring constants by the equations $k_x = k_y = m \omega_{orb}^2$.  As mentioned above, we take $k_z \rightarrow \infty$ since confinement is strong along the growth direction.  The base temperature is $25\,$mK, while the electron temperature is about $150\,$mK, the value we use for the calculations. 

Another parameter needed as input to the theory is the conductivity $\sigma$ of the Au gates.  This was not measured in this device, but under similar growth conditions for Au films a value of $\sigma = 2\times 10^8\,$S\,m$^{-1}$ was obtained at the temperatures of the experiment \cite{de1987temperature}. We should regard this as a probably somewhat high order-of-magnitude estimate of $\sigma$ in the actual device, and we used a range of values for $\sigma$. We take the gate  thickness as $w= 25\,$nm, and the distance from the qubit to the gates $d = 137\,$nm.  The qubit operating frequency is $\omega_{op} = 2\pi \times 12.9\,$GHz $= 8.11 \times 10^{10}\,\mathrm{s}^{-1}$. Two micromagnets made of cobalt are defined on top of the gates, approximately $162\,$nm above the qubit.   In the experiments reported in Ref.~\cite{Kawakami:2014p666} the applied field $B_0$ was in the $x$-direction.  This paper concerns what happens if this direction is varied (but with the caveats discussed in Sec.~\ref{subsec:magnet})
.


\section*{Data Availability}
Authors can confirm that all relevant data are included in the article and/or its supporting files.

\section*{Code Availability}
The code that produced the figures in the text is available from the corresponding author upon reasonable request.


\begin{acknowledgements}
We thank M.A. Eriksson, M. Friesen, L. Tom, and S.N. Coppersmith for useful discussions, and L.M.K. Vandersypen and E. Kawakami for detailed information about the device used in the experiment described in Ref. \cite{Kawakami:2014p666}.  This research was sponsored in part by the Army Research Office (ARO) under Grant Number W911NF-17-1-0274. The views and conclusions contained in this document are those of the authors and should not be interpreted as representing the official policies, either expressed or implied, of the Army Research Office (ARO), or the U.S. Government. The U.S. Government is authorized to reproduce and distribute reprints for Government purposes notwithstanding any copyright notation herein.
\end{acknowledgements}


\section*{Author Contribution}
R.J.J. conceived and supervised the study. Y.C. modeled the problem and obtained the simulated results. All authors discussed the results and prepared the manuscript.

\section*{Competing Interests}
The authors declare that there are no competing interests.

\bibliographystyle{naturemag}
\bibliography{main}

\begin{thebibliography}{10}
\expandafter\ifx\csname url\endcsname\relax
  \def\url#1{\texttt{#1}}\fi
\expandafter\ifx\csname urlprefix\endcsname\relax\def\urlprefix{URL }\fi
\providecommand{\bibinfo}[2]{#2}
\providecommand{\eprint}[2][]{\url{#2}}

\bibitem{Zwanenburg:2013p961}
\bibinfo{author}{Zwanenburg, F.~A.} \emph{et~al.}
\newblock \bibinfo{title}{Silicon quantum electronics}.
\newblock \emph{\bibinfo{journal}{Rev. Mod. Phys.}}
  \textbf{\bibinfo{volume}{85}}, \bibinfo{pages}{961} (\bibinfo{year}{2013}).

\bibitem{Yoneda:2018p102}
\bibinfo{author}{Yoneda, J.} \emph{et~al.}
\newblock \bibinfo{title}{{A quantum-dot spin qubit with coherence limited by
  charge noise and fidelity higher than 99.9\%}}.
\newblock \emph{\bibinfo{journal}{Nat. Nanotechnol.}}
  \textbf{\bibinfo{volume}{13}}, \bibinfo{pages}{102--106}
  (\bibinfo{year}{2018}).

\bibitem{Mi:2018p599}
\bibinfo{author}{Mi, X.} \emph{et~al.}
\newblock \bibinfo{title}{{A coherent spin-photon interface in silicon}}.
\newblock \emph{\bibinfo{journal}{Nature}} \textbf{\bibinfo{volume}{555}},
  \bibinfo{pages}{599--603} (\bibinfo{year}{2018}).

\bibitem{Watson:2018p633}
\bibinfo{author}{Watson, T.~F.} \emph{et~al.}
\newblock \bibinfo{title}{A programmable two-qubit quantum processor in
  silicon}.
\newblock \emph{\bibinfo{journal}{Nature}} \textbf{\bibinfo{volume}{555}},
  \bibinfo{pages}{633--637} (\bibinfo{year}{2018}).

\bibitem{petit2020universal}
\bibinfo{author}{Petit, L.} \emph{et~al.}
\newblock \bibinfo{title}{Universal quantum logic in hot silicon qubits}.
\newblock \emph{\bibinfo{journal}{Nature}} \textbf{\bibinfo{volume}{580}},
  \bibinfo{pages}{355--359} (\bibinfo{year}{2020}).

\bibitem{neyens2019measurements}
\bibinfo{author}{Neyens, S.~F.} \emph{et~al.}
\newblock \bibinfo{title}{Measurements of capacitive coupling within a
  quadruple-quantum-dot array}.
\newblock \emph{\bibinfo{journal}{Phys. Rev. Appl.}}
  \textbf{\bibinfo{volume}{12}}, \bibinfo{pages}{064049}
  (\bibinfo{year}{2019}).

\bibitem{Coish:2009p2203}
\bibinfo{author}{Coish, W.~A.} \& \bibinfo{author}{Baugh, J.}
\newblock \bibinfo{title}{Nuclear spins in nanostructures}.
\newblock \emph{\bibinfo{journal}{Phys. Status Solidi (b)}}
  \textbf{\bibinfo{volume}{246}}, \bibinfo{pages}{2203--2215}
  (\bibinfo{year}{2009}).

\bibitem{Beaudoin:2013p085320}
\bibinfo{author}{Beaudoin, F.} \& \bibinfo{author}{Coish, W.~A.}
\newblock \bibinfo{title}{Enhanced hyperfine-induced spin dephasing in a
  magnetic-field gradient}.
\newblock \emph{\bibinfo{journal}{Phys. Rev. B}} \textbf{\bibinfo{volume}{88}},
  \bibinfo{pages}{085320} (\bibinfo{year}{2013}).

\bibitem{LangsjoenPRA2012}
\bibinfo{author}{Langsjoen, L.~S.}, \bibinfo{author}{Poudel, A.},
  \bibinfo{author}{Vavilov, M.~G.} \& \bibinfo{author}{Joynt, R.}
\newblock \bibinfo{title}{Qubit relaxation from evanescent-wave {J}ohnson
  noise}.
\newblock \emph{\bibinfo{journal}{Phys. Rev. A}} \textbf{\bibinfo{volume}{86}},
  \bibinfo{pages}{010301} (\bibinfo{year}{2012}).

\bibitem{premakumar2017}
\bibinfo{author}{Premakumar, V.}, \bibinfo{author}{Vavilov, M.} \&
  \bibinfo{author}{Joynt, R.}
\newblock \bibinfo{title}{Evanescent-wave {J}ohnson noise in small devices}.
\newblock \emph{\bibinfo{journal}{Quantum Sci. Tech.}}
  \textbf{\bibinfo{volume}{3}}, \bibinfo{pages}{105001} (\bibinfo{year}{2017}).

\bibitem{Veldhorst:2014p981}
\bibinfo{author}{Veldhorst, M.} \emph{et~al.}
\newblock \bibinfo{title}{An addressable quantum dot qubit with fault-tolerant
  control-fidelity}.
\newblock \emph{\bibinfo{journal}{Nat. Nanotechnol.}}
  \textbf{\bibinfo{volume}{9}}, \bibinfo{pages}{981--985}
  (\bibinfo{year}{2014}).

\bibitem{eng2015isotopically}
\bibinfo{author}{Eng, K.} \emph{et~al.}
\newblock \bibinfo{title}{Isotopically enhanced triple-quantum-dot qubit}.
\newblock \emph{\bibinfo{journal}{Sci. Adv.}} \textbf{\bibinfo{volume}{1}},
  \bibinfo{pages}{e1500214} (\bibinfo{year}{2015}).

\bibitem{connors2019low}
\bibinfo{author}{Connors, E.~J.}, \bibinfo{author}{Nelson, J.},
  \bibinfo{author}{Qiao, H.}, \bibinfo{author}{Edge, L.~F.} \&
  \bibinfo{author}{Nichol, J.~M.}
\newblock \bibinfo{title}{Low-frequency charge noise in {S}i/{S}i{G}e quantum
  dots}.
\newblock \emph{\bibinfo{journal}{Phys. Rev. B}}
  \textbf{\bibinfo{volume}{100}}, \bibinfo{pages}{165305}
  (\bibinfo{year}{2019}).

\bibitem{khaetskii2001spin}
\bibinfo{author}{Khaetskii, A.~V.} \& \bibinfo{author}{Nazarov, Y.~V.}
\newblock \bibinfo{title}{Spin-flip transitions between {Z}eeman sublevels in
  semiconductor quantum dots}.
\newblock \emph{\bibinfo{journal}{Phys. Rev. B}} \textbf{\bibinfo{volume}{64}},
  \bibinfo{pages}{125316} (\bibinfo{year}{2001}).

\bibitem{Tahan:2002p035314}
\bibinfo{author}{Tahan, C.}, \bibinfo{author}{Friesen, M.} \&
  \bibinfo{author}{Joynt, R.}
\newblock \bibinfo{title}{Decoherence of electron spin qubits in {S}i-based
  quantum computers}.
\newblock \emph{\bibinfo{journal}{Phys. Rev. B}} \textbf{\bibinfo{volume}{66}},
  \bibinfo{pages}{035314} (\bibinfo{year}{2002}).

\bibitem{Amasha:2008p2332}
\bibinfo{author}{Amasha, S.} \emph{et~al.}
\newblock \bibinfo{title}{Electrical control of spin relaxation in a quantum
  dot}.
\newblock \emph{\bibinfo{journal}{Phys. Rev. Lett.}}
  \textbf{\bibinfo{volume}{100}}, \bibinfo{pages}{046803}
  (\bibinfo{year}{2008}).

\bibitem{Koh:2012p250503}
\bibinfo{author}{Koh, T.~S.}, \bibinfo{author}{Gamble, J.~K.},
  \bibinfo{author}{Friesen, M.}, \bibinfo{author}{Eriksson, M.~A.} \&
  \bibinfo{author}{Coppersmith, S.~N.}
\newblock \bibinfo{title}{Pulse-gated quantum dot hybrid qubit}.
\newblock \emph{\bibinfo{journal}{Phys. Rev. Lett.}}
  \textbf{\bibinfo{volume}{109}}, \bibinfo{pages}{250503}
  (\bibinfo{year}{2012}).

\bibitem{Johnson:2005p483}
\bibinfo{author}{Johnson, A.~C.}, \bibinfo{author}{Petta, J.~R.},
  \bibinfo{author}{Marcus, C.~M.}, \bibinfo{author}{Hanson, M.~P.} \&
  \bibinfo{author}{Gossard, A.~C.}
\newblock \bibinfo{title}{Singlet-triplet spin blockade and charge sensing in a
  few-electron double quantum dot}.
\newblock \emph{\bibinfo{journal}{Phys. Rev. B}} \textbf{\bibinfo{volume}{72}},
  \bibinfo{pages}{165308} (\bibinfo{year}{2005}).

\bibitem{seedhouse2021pauli}
\bibinfo{author}{Seedhouse, A.~E.} \emph{et~al.}
\newblock \bibinfo{title}{Pauli blockade in silicon quantum dots with
  spin-orbit control}.
\newblock \emph{\bibinfo{journal}{PRX Quantum}} \textbf{\bibinfo{volume}{2}},
  \bibinfo{pages}{010303} (\bibinfo{year}{2021}).

\bibitem{camenzind2018hyperfine}
\bibinfo{author}{Camenzind, L.~C.} \emph{et~al.}
\newblock \bibinfo{title}{Hyperfine-phonon spin relaxation in a single-electron
  {G}a{A}s quantum dot}.
\newblock \emph{\bibinfo{journal}{Nat. Commun.}} \textbf{\bibinfo{volume}{9}},
  \bibinfo{pages}{3454} (\bibinfo{year}{2018}).

\bibitem{Raith_PRL2012}
\bibinfo{author}{Raith, M.}, \bibinfo{author}{Stano, P.},
  \bibinfo{author}{Baruffa, F.} \& \bibinfo{author}{Fabian, J.}
\newblock \bibinfo{title}{Theory of spin relaxation in two-electron lateral
  coupled quantum dots}.
\newblock \emph{\bibinfo{journal}{Phys. Rev. Lett.}}
  \textbf{\bibinfo{volume}{108}}, \bibinfo{pages}{246602}
  (\bibinfo{year}{2012}).

\bibitem{xiao2010measurement}
\bibinfo{author}{Xiao, M.}, \bibinfo{author}{House, M.} \&
  \bibinfo{author}{Jiang, H.~W.}
\newblock \bibinfo{title}{Measurement of the spin relaxation time of single
  electrons in a silicon metal-oxide-semiconductor-based quantum dot}.
\newblock \emph{\bibinfo{journal}{Phys. Rev. Lett.}}
  \textbf{\bibinfo{volume}{104}}, \bibinfo{pages}{096801}
  (\bibinfo{year}{2010}).

\bibitem{Hao:2014p3860}
\bibinfo{author}{Hao, X.}, \bibinfo{author}{Ruskov, R.}, \bibinfo{author}{Xiao,
  M.}, \bibinfo{author}{Tahan, C.} \& \bibinfo{author}{Jiang, H.~W.}
\newblock \bibinfo{title}{Electron spin resonance and spin-valley physics in a
  silicon double quantum dot}.
\newblock \emph{\bibinfo{journal}{Nat. Commun.}} \textbf{\bibinfo{volume}{5}},
  \bibinfo{pages}{3860} (\bibinfo{year}{2014}).

\bibitem{ferdous2018interface}
\bibinfo{author}{Ferdous, R.} \emph{et~al.}
\newblock \bibinfo{title}{Interface-induced spin-orbit interaction in silicon
  quantum dots and prospects for scalability}.
\newblock \emph{\bibinfo{journal}{Phys. Rev. B}} \textbf{\bibinfo{volume}{97}},
  \bibinfo{pages}{241401} (\bibinfo{year}{2018}).

\bibitem{sigillito2017all}
\bibinfo{author}{Sigillito, A.~J.}, \bibinfo{author}{Tyryshkin, A.~M.},
  \bibinfo{author}{Schenkel, T.}, \bibinfo{author}{Houck, A.~A.} \&
  \bibinfo{author}{Lyon, S.~A.}
\newblock \bibinfo{title}{All-electric control of donor nuclear spin qubits in
  silicon}.
\newblock \emph{\bibinfo{journal}{Nat. Nanotechnol.}}
  \textbf{\bibinfo{volume}{12}}, \bibinfo{pages}{958--962}
  (\bibinfo{year}{2017}).

\bibitem{zhang2021controlling}
\bibinfo{author}{Zhang, X.} \emph{et~al.}
\newblock \bibinfo{title}{Controlling synthetic spin-orbit coupling in a
  silicon quantum dot with magnetic field}.
\newblock \emph{\bibinfo{journal}{Phys. Rev. Appl.}}
  \textbf{\bibinfo{volume}{15}}, \bibinfo{pages}{044042}
  (\bibinfo{year}{2021}).

\bibitem{crippa2018electrical}
\bibinfo{author}{Crippa, A.} \emph{et~al.}
\newblock \bibinfo{title}{Electrical spin driving by g-matrix modulation in
  spin-orbit qubits}.
\newblock \emph{\bibinfo{journal}{Phys. Rev. Lett.}}
  \textbf{\bibinfo{volume}{120}}, \bibinfo{pages}{137702}
  (\bibinfo{year}{2018}).

\bibitem{tanttu2019controlling}
\bibinfo{author}{Tanttu, T.} \emph{et~al.}
\newblock \bibinfo{title}{Controlling spin-orbit interactions in silicon
  quantum dots using magnetic field direction}.
\newblock \emph{\bibinfo{journal}{Phys. Rev. X}} \textbf{\bibinfo{volume}{9}},
  \bibinfo{pages}{021028} (\bibinfo{year}{2019}).

\bibitem{Marx:2020}
\bibinfo{author}{Marx, M.} \emph{et~al.}
\newblock \bibinfo{title}{Spin orbit field in a physically defined p type mos
  silicon double quantum dot}.
\newblock \emph{\bibinfo{journal}{Preprint at
  https://arxiv.org/abs/2003.07079}}  (\bibinfo{year}{2020}).

\bibitem{Scarlino:2015p106802}
\bibinfo{author}{Scarlino, P.} \emph{et~al.}
\newblock \bibinfo{title}{Second-harmonic coherent driving of a spin qubit in a
  {S}i/{S}i{G}e quantum dot}.
\newblock \emph{\bibinfo{journal}{Phys. Rev. Lett.}}
  \textbf{\bibinfo{volume}{115}}, \bibinfo{pages}{106802}
  (\bibinfo{year}{2015}).

\bibitem{takeda2018optimized}
\bibinfo{author}{Takeda, K.} \emph{et~al.}
\newblock \bibinfo{title}{Optimized electrical control of a {S}i/{S}i{G}e spin
  qubit in the presence of an induced frequency shift}.
\newblock \emph{\bibinfo{journal}{npj Quantum Inf.}}
  \textbf{\bibinfo{volume}{4}}, \bibinfo{pages}{54} (\bibinfo{year}{2018}).

\bibitem{tokura2006coherent}
\bibinfo{author}{Tokura, Y.}, \bibinfo{author}{van~der Wiel, W.~G.},
  \bibinfo{author}{Obata, T.} \& \bibinfo{author}{Tarucha, S.}
\newblock \bibinfo{title}{Coherent single electron spin control in a slanting
  zeeman field}.
\newblock \emph{\bibinfo{journal}{Phys. Rev. Lett.}}
  \textbf{\bibinfo{volume}{96}}, \bibinfo{pages}{047202}
  (\bibinfo{year}{2006}).

\bibitem{Kawakami:2014p666}
\bibinfo{author}{Kawakami, E.} \emph{et~al.}
\newblock \bibinfo{title}{Electrical control of a long-lived spin qubit in a
  {Si/SiGe} quantum dot}.
\newblock \emph{\bibinfo{journal}{Nat. Nanotechnol.}}
  \textbf{\bibinfo{volume}{9}}, \bibinfo{pages}{666--670}
  (\bibinfo{year}{2014}).

\bibitem{neumann2015}
\bibinfo{author}{Neumann, R.} \& \bibinfo{author}{Schreiber, L.}
\newblock \bibinfo{title}{Simulation of micro-magnet stray field dynamics for
  spin qubit manipulation}.
\newblock \emph{\bibinfo{journal}{J. Appl. Phys.}}
  \textbf{\bibinfo{volume}{117}}, \bibinfo{pages}{193903}
  (\bibinfo{year}{2015}).

\bibitem{chesi2014single}
\bibinfo{author}{Chesi, S.} \emph{et~al.}
\newblock \bibinfo{title}{Single-spin manipulation in a double quantum dot in
  the field of a micromagnet}.
\newblock \emph{\bibinfo{journal}{Phys. Rev. B}} \textbf{\bibinfo{volume}{90}},
  \bibinfo{pages}{235311} (\bibinfo{year}{2014}).

\bibitem{ferdous2018valley}
\bibinfo{author}{Ferdous, R.} \emph{et~al.}
\newblock \bibinfo{title}{Valley dependent anisotropic spin splitting in
  silicon quantum dots}.
\newblock \emph{\bibinfo{journal}{npj Quantum Inf.}}
  \textbf{\bibinfo{volume}{4}}, \bibinfo{pages}{26} (\bibinfo{year}{2018}).

\bibitem{de1998chemical}
\bibinfo{author}{De~Campos, M.} \emph{et~al.}
\newblock \bibinfo{title}{Chemical composition and coercivity of {S}m{C}o$_5$
  magnets}.
\newblock \emph{\bibinfo{journal}{J. Appl. Phys}}
  \textbf{\bibinfo{volume}{84}}, \bibinfo{pages}{368--373}
  (\bibinfo{year}{1998}).

\bibitem{henkel1999}
\bibinfo{author}{Henkel, C.} \& \bibinfo{author}{Wilkens, M.}
\newblock \bibinfo{title}{Heating of trapped atoms near thermal surfaces}.
\newblock \emph{\bibinfo{journal}{Europhys. Lett.}}
  \textbf{\bibinfo{volume}{47}}, \bibinfo{pages}{414--420}
  (\bibinfo{year}{1999}).

\bibitem{langsjoen2014}
\bibinfo{author}{Langsjoen, L.}, \bibinfo{author}{Poudel, A.},
  \bibinfo{author}{Vavilov, M.} \& \bibinfo{author}{Joynt, R.}
\newblock \bibinfo{title}{Electromagnetic fluctuations near metallic thin
  films}.
\newblock \emph{\bibinfo{journal}{Phys. Rev. B}} \textbf{\bibinfo{volume}{89}},
  \bibinfo{pages}{115401} (\bibinfo{year}{2014}).

\bibitem{anderson1972anomalous}
\bibinfo{author}{Anderson, P.~W.}, \bibinfo{author}{Halperin, B.~I.} \&
  \bibinfo{author}{Varma, C.~M.}
\newblock \bibinfo{title}{Anomalous low-temperature thermal properties of
  glasses and spin glasses}.
\newblock \emph{\bibinfo{journal}{Philos. Mag.}} \textbf{\bibinfo{volume}{25}},
  \bibinfo{pages}{1--9} (\bibinfo{year}{1972}).

\bibitem{phillips1972tunneling}
\bibinfo{author}{Phillips, W.~A.}
\newblock \bibinfo{title}{Tunneling states in amorphous solids}.
\newblock \emph{\bibinfo{journal}{J. Low Temp. Phys.}}
  \textbf{\bibinfo{volume}{7}}, \bibinfo{pages}{351--360}
  (\bibinfo{year}{1972}).

\bibitem{mcwhorter1957semiconductor}
\bibinfo{author}{McWhorter, A.} \emph{et~al.}
\newblock \bibinfo{title}{Semiconductor surface physics}.
\newblock \emph{\bibinfo{journal}{University of Pennsylvania Press,
  Philadelphia, PA}} \textbf{\bibinfo{volume}{207}} (\bibinfo{year}{1957}).

\bibitem{Takeda:2016p1600694}
\bibinfo{author}{Takeda, K.} \emph{et~al.}
\newblock \bibinfo{title}{A fault-tolerant addressable spin qubit in a natural
  silicon quantum dot}.
\newblock \emph{\bibinfo{journal}{Sci. Adv.}} \textbf{\bibinfo{volume}{2}},
  \bibinfo{pages}{e1600694} (\bibinfo{year}{2016}).

\bibitem{ahn2021microscopic}
\bibinfo{author}{Ahn, S.}, \bibinfo{author}{Das~Sarma, S.} \&
  \bibinfo{author}{Kestner, J.~P.}
\newblock \bibinfo{title}{Microscopic bath effects on noise spectra in
  semiconductor quantum dot qubits}.
\newblock \emph{\bibinfo{journal}{Phys. Rev. B}}
  \textbf{\bibinfo{volume}{103}}, \bibinfo{pages}{L041304}
  (\bibinfo{year}{2021}).

\bibitem{zhang2017selective}
\bibinfo{author}{Zhang, H.}, \bibinfo{author}{Arai, K.},
  \bibinfo{author}{Belthangady, C.}, \bibinfo{author}{Jaskula, J.-C.} \&
  \bibinfo{author}{Walsworth, R.~L.}
\newblock \bibinfo{title}{Selective addressing of solid-state spins at the
  nanoscale via magnetic resonance frequency encoding}.
\newblock \emph{\bibinfo{journal}{npj Quantum Inf.}}
  \textbf{\bibinfo{volume}{3}}, \bibinfo{pages}{31} (\bibinfo{year}{2017}).

\bibitem{zimmermann2020selective}
\bibinfo{author}{Zimmermann, J.} \emph{et~al.}
\newblock \bibinfo{title}{Selective noise resistant gate}.
\newblock \emph{\bibinfo{journal}{Phys.~Rev.~B}}
  \textbf{\bibinfo{volume}{102}}, \bibinfo{pages}{245408}
  (\bibinfo{year}{2020}).

\bibitem{levinshtein2001properties}
\bibinfo{author}{Levinshtein, M.~E.}, \bibinfo{author}{Rumyantsev, S.~L.} \&
  \bibinfo{author}{Shur, M.~S.}
\newblock \emph{\bibinfo{title}{Properties of Advanced Semiconductor Materials:
  GaN, AIN, InN, BN, SiC, SiGe}} (\bibinfo{publisher}{John Wiley \& Sons},
  \bibinfo{year}{2001}).

\bibitem{de1987temperature}
\bibinfo{author}{De~Vries, J.}
\newblock \bibinfo{title}{Temperature-dependent resistivity measurements on
  polycrystalline $\mathrm{Si}\mathrm{O}_2$-covered thin gold films}.
\newblock \emph{\bibinfo{journal}{Thin Solid Films}}
  \textbf{\bibinfo{volume}{150}}, \bibinfo{pages}{201--208}
  (\bibinfo{year}{1987}).

\bibitem{kawakamithesis}
\bibinfo{author}{Kawakami, E.}
\newblock \emph{\bibinfo{title}{Characterization of an electron spin qubit in a
  Si/SiGe quantum dot}}.
\newblock Ph.D. thesis, \bibinfo{school}{Delft University of Technology}
  (\bibinfo{year}{2016}).

\end{thebibliography}

%

\onecolumngrid
\appendix

\clearpage
\newpage

\section*{Supplementary Note I: Relaxation time due to Evanescent-wave Johnson noise}

This section gives further details for the calculation of $T_1$ of a spin qubit affected by evanescent-wave Johnson noise (EWJN). There is both direct magnetic noise $\vec{B}(t)$ and indirect magnetic noise $\vec{B}^{(E)}$ due to the electrically-induced motion of the qubit in the magnetic field gradient.  

The correlation function of total noise field $\vec{B}^{\rm{(eff)}}$ can be expanded as
\begin{equation}
\begin{aligned}
\langle B_i^{\rm{(eff)}}  B_j^{\rm{(eff)}} \rangle &= \langle (B_i + B_i^{(E)})   (B_j + B_j^{(E)})  \rangle \\
&=  \langle B_i B_j  \rangle +  \langle B_i   B_j^{(E)}  \rangle +  \langle  B_i^{(E)} B_j   \rangle + \langle B_i^{(E)}  B_j^{(E)} \rangle \\
&= \langle B_i B_j  \rangle + \sum_{mn} \left[  \frac{q^2}{4k_m k_n} \frac{\partial B_{i}}{\partial x_m} \frac{\partial B_{j}}{\partial x_n} \langle  E_m  E_n \rangle \right. \\
&\quad \left. + \frac{q}{2k_m} \frac{\partial B_{j}}{\partial x_m} \langle B_i  E_m \rangle + \frac{q}{2k_n} \frac{\partial B_{i}}{\partial x_n} \langle  E_n B_j \rangle \right].
\end{aligned}
\end{equation}
Using Eqs. (6) and (7) of the main text for the dynamics of $ B_i^{(E)}(t)$ and assuming that there is no correlation between $ B_i(t)$ and $ B_i^{(E)}(t)$, we arrive at the correlation functions for the effective magnetic field.

The correlation functions become
\begin{equation}
\begin{aligned}
\langle B_x^{\rm{(eff)}}  B_x^{\rm{(eff)}} \rangle
&=  \langle B_x B_x \rangle + (\frac{q}{2m\omega_{orb}^2})^2 \left\{ (\frac{\partial B_{x}}{\partial x})^2 \langle  E_x  E_x \rangle + (\frac{\partial B_{x}}{\partial y})^2 \langle  E_y  E_y \rangle \right\} \\
&= \left[ \frac{\mu_0 \hbar \omega_{op} \sigma  w}{16 d^2} + (\frac{q}{2m \omega_{orb}^2})^2 \frac{\hbar \omega_{op} \epsilon_d \epsilon_0}{4 \sigma d^3} \left\{  (\frac{\partial B_{x}}{\partial x})^2 +  (\frac{\partial B_{x}}{\partial y})^2  \right\} \right] \coth \frac{\hbar \omega_{op}}{2 k_B T}, \\
\langle B_y^{\rm{(eff)}}  B_y^{\rm{(eff)}} \rangle
&=  \langle B_y B_y \rangle + (\frac{q}{2m\omega_{orb}^2})^2 \left\{ (\frac{\partial B_{y}}{\partial x})^2 \langle  E_x  E_x \rangle + (\frac{\partial B_{y}}{\partial y})^2 \langle  E_y  E_y \rangle \right\} \\
&= \left[ \frac{\mu_0 \hbar \omega_{op} \sigma  w}{16 d^2} + (\frac{q}{2m \omega_{orb}^2})^2 \frac{\hbar \omega_{op} \epsilon_d \epsilon_0}{4 \sigma d^3} \left\{  (\frac{\partial B_{y}}{\partial x})^2 +  (\frac{\partial B_{y}}{\partial y})^2  \right\} \right] \coth \frac{\hbar \omega_{op}}{2 k_B T}, \\
\langle B_z^{\rm{(eff)}}  B_z^{\rm{(eff)}} \rangle
&=  \langle B_z B_z \rangle + (\frac{q}{2m\omega_{orb}^2})^2 \left\{ (\frac{\partial B_{z}}{\partial x})^2 \langle  E_x  E_x \rangle + (\frac{\partial B_{z}}{\partial y})^2 \langle  E_y  E_y \rangle \right\} \\
&= \left[ \frac{\mu_0 \hbar \omega_{op} \sigma  w}{8 d^2} + (\frac{q}{2m \omega_{orb}^2})^2 \frac{\hbar \omega_{op} \epsilon_d \epsilon_0}{4 \sigma d^3} \left\{  (\frac{\partial B_{z}}{\partial x})^2 +  (\frac{\partial B_{z}}{\partial y})^2  \right\} \right] \coth \frac{\hbar \omega_{op}}{2 k_B T},
\end{aligned}
\end{equation}
while the terms $\langle B_i^{\rm{(eff)}}  B_j^{\rm{(eff)}} \rangle$ with $i \neq j$ are
\begin{equation}
\langle B_i^{\rm{(eff)}}  B_j^{\rm{(eff)}} \rangle = (\frac{q}{2m \omega_{orb}^2})^2 \frac{\hbar \omega_{op} \epsilon_d \epsilon_0}{4 \sigma d^3} \left\{  \frac{\partial B_{i}}{\partial x} \frac{\partial B_{j}}{\partial x} +  \frac{\partial B_{i}}{\partial y} \frac{\partial B_{j}}{\partial y}  \right\} \coth \frac{\hbar \omega_{op}}{2 k_B T}.
\end{equation}

Substituting Eqs. (8) and (9) for the EWJN into equation of this kind, and applying Eq. (3) to rotate the results into a general direction, we arrive at Eqs. (18) and (19) from which the results in Fig. 2 are calculated.

Eqs. (8) and (9) themselves are the half-space local (point qubit) correlation functions for EWJN \cite{premakumar2017}. They are valid only in the $d \ll \delta$ regime.  The distance $d$ between the qubit and accumulation gates is about $137\,$nm while skin depth $\delta = \sqrt{2/ \mu_0 \sigma \omega_{op}} = 313\,$nm with the conductivity of gates estimated very roughly as $\sigma = 2 \times 10^8\,$S/m \cite{de1987temperature} and the operating frequency $\omega_{op} = 2\pi \times 12.9\,$GHz $= 8.11 \times 10^{10}\,\rm{s}^{-1}$. Thus the device parameters satisfy the $d \ll \delta$ requirement.  

The dephasing time $T_{\phi}$ is calculated in a similar fashion, with the main difference being that it only depends on a single correlation function.

The Gaussian approximation for $T_{\phi}$ should be valid for EWJN since the noise comes from a large number of modes in the solid.  The decay in the off-diagonal components of the density matrix is initially Gaussian in time and then crosses over to exponential.  The detailed time dependence is given by $\exp[-\Gamma(t)]$ with $\Gamma(t)$ given by Eq. (4) in the main text.  $T_{\phi}$ is then obtained by solving the transcendental equation $\Gamma(T_{\phi})=1$.  When rotated into the $(\theta,\phi)$ direction we obtain Eq. (5).

\section*{Supplementary Note II: Correlation function of Charge noise}

In this section we give further details of calculations in the Anisotropy of $T_2$ section in the main text.
The relaxation time due to charge noise can be obtained by following the same way as that of EWJN with Eqs.(2) of the main text. The difference is that the correlation function of effective noise field now consists only of the electric noise part such that
\begin{equation}
\label{effbb}
\langle B_i^{\rm{(eff)}}  B_j^{\rm{(eff)}} \rangle = \sum_{mn}  \frac{q^2}{4k_m k_n} \frac{\partial B_{i}}{\partial x_m} \frac{\partial B_{j}}{\partial x_n} \langle  E_m  E_n \rangle.
\end{equation}
The electric fields from a point dipole are given as
\begin{equation}
\begin{aligned}
 E_i(\vec{r},t) &= \frac{1}{4 \pi \epsilon_0} \sum_k  \frac{3 R'_k p_k(t) R'_i - R'_k R'_k p_i(t)}{|\vec{R}'|^5}, \\
 E_j(\vec{r},0) &= \frac{1}{4 \pi \epsilon_0} \sum_m  \frac{3 R'_m p_m(0) R'_j - R'_m R'_m p_j(0)}{|\vec{R}'|^5}
 \end{aligned}
 \end{equation}
 where $\vec{R}' = \vec{r} - \vec{r}'$ is the displacement vector from the position of a dipole $\vec{r}'$ to that of a qubit $\vec{r}$. The dipoles are assumed to be statistically independent, so their correlation function is
 \begin{equation}
     \langle p_i(\vec{r}',t) p_j(\vec{r}',0) \rangle =
      \delta_{mod} \, p_0^2 \, g(t).
 \end{equation}
 where model dependent factor $\delta_{mod} = \delta_{ij}/3$ for random dipole model and $\delta_{mod} = \delta_{iz} \delta_{jz}$ for trap model. The time correlation function is assumed to be exponential: $g(t) = e^{-|t|/\tau}$ where $\tau$ is the characteristic time of the dipoles. This corresponds to random telegraph noise, but the precise model for the time correlations is not that important, since the anisotropy maps depend only on ratios of specific averages of the correlation functions.
 
 Combining all the equations above, the electric field correlation function is
 \begin{equation}
 \begin{aligned}
     \langle E_i(\vec{r},t) E_j(\vec{r},0) \rangle = \frac{1}{16 \pi^2 \epsilon_0^2} \sum_{km} |\vec{R}'|^{-10} &\langle\, 
     9 R'_k R'_m R'_i R'_j p_k (t) p_m (0)
     - 3 R'_m R'_j R'_k R'_k p_i (t) p_m (0)  \\
     &- 3 R'_k R'_i R'_m R'_m p_j (t) p_k (0)
     + R'_k R'_k R'_m R'_m p_i (t) p_j (0) \,\rangle,
\end{aligned}
\end{equation}
which simplifies to
 \begin{equation}
 \label{dipoleee}
  \langle E_i(\vec{r},t)  E_j(\vec{r},0) \rangle = 
  \frac{1}{48 \pi^2 \epsilon_0^2} \rho_v p_0^2 \, g(\omega) \int d^3r'
  \frac{3 R'_i R'_j + \delta_{ij} |\vec{R}'|^2}{|\vec{R}'|^8}
 \end{equation}
 for the random dipole model and
  \begin{equation}
  \label{trapee}
  \begin{aligned}
  \langle E_i(\vec{r},t)  E_j(\vec{r},0) \rangle &= 
  \frac{1}{16 \pi^2 \epsilon_0^2} \rho_a p_0^2 \, g(\omega) \\
  &\times \int d^2r'
  \frac{9 R_z^{\prime 2} R'_i R'_j - 3R'_z R'_j |\vec{R}'|^2 \delta_{iz} - 3R'_z R'_i |\vec{R}'|^2 \delta_{jz} + |\vec{R}'|^4 \delta_{iz}\delta_{jz} }{|\vec{R}'|^{10}}
  \end{aligned}
 \end{equation}
 for the trap model.
 Here $\rho_v$ is the volume density of dipoles and $\rho_a$ is the areal density of traps, which are assumed to be uniformly distributed in the region of integration and used as fitting parameters to make $T_2^{(x)} = 840\,$ns. $g(\omega) = 2\tau/(1+(\omega_{op} \tau)^2)$ is the Fourier transform of $g(t)$ and $\tau$ is taken as the inverse of maximum attempt frequency in \cite{connors2019low}. The integration region for the uniformly distributed random dipole model (UD) is the aluminum oxide layer which is infinite in $xy$ plane, about $37\,$nm above the qubit, and whose thickness is about $l = 100\,$nm, namely $\rho \in [0,\infty),\; \phi \in [0,2\pi),\; z \in [37, 137]\,$nm with cylindrical coordinate system. That for uniformly distributed trap model (UT) is the interface between the oxide layer and accumulation gates that is about $d = 137\,$nm above the qubit, \textit{i.e.} $\rho \in [0,\infty),\; \phi \in [-\pi/4,5\pi/4]$, at $z = d$. The domain of $\phi \in [-\pi/4,5\pi/4]$ takes into account the actual gate geometry of the device of Kawakami \textit{et al.} \cite{kawakamithesis}.
 
\section*{Supplementary Note III: Relaxation time due to Charge noise}

The relaxation rate with applied field in $(\theta, \phi)$ direction is then
\begin{equation}
\frac{1}{ T_1 (\theta, \phi) } = \sum_{ij} Q^{(1)}_{ij}  \left[ \gamma_{1x} \frac{\partial B_{i} }{\partial x} \frac{\partial B_{j} }{\partial x} + \gamma_{1y} \frac{\partial B_{i} }{\partial y} \frac{\partial B_{j} }{\partial y} \right]
\end{equation}
where $\gamma_{1x}$ and $\gamma_{1y}$ are the prefactors related to the gradient in $x$- and $y$-direction respectively, and determined by experimental parameters as follows: with the UD model
\begin{equation}
    \gamma_{1x} = \gamma_{1y} = \; (\frac{\mu_B}{\hbar})^2 (\frac{q}{2m w_{orb}^2})^2 \frac{\rho_v p_0^2}{576 \pi \epsilon_0^2} \left( \frac{1}{l^3} - \frac{1}{d^3} \right) \frac{2\tau}{1+\omega_{op}\tau},
\end{equation}
and with the UT model
\begin{equation}
\begin{aligned}
    \gamma_{1x} = \; &(\frac{\mu_B}{\hbar})^2 (\frac{q}{2m w_{orb}^2})^2  \frac{(9\pi+6) \rho_a p_0^2}{512 \pi^2 \epsilon_0^2 d^4} \frac{2\tau}{1+\omega_{op}\tau}, \\
    \gamma_{1y} = \; &(\frac{\mu_B}{\hbar})^2 (\frac{q}{2m w_{orb}^2})^2  \frac{(9\pi-6) \rho_a p_0^2}{512 \pi^2 \epsilon_0^2 d^4} \frac{2\tau}{1+\omega_{op}\tau}.
\end{aligned}
\end{equation}
The simulation results of $T_1^{(x)}$ are $6.50 \times 10^9\,$s for the UD model and $1.72 \times 10^{10}\,$s for the UT model with the fitted volume and areal densities respectively.

\section*{Supplementary Note IV: Dephasing time due to Charge noise}

Using the above Eqs. \ref{effbb}, \ref{dipoleee}, and \ref{trapee}, $\Gamma(t)$ for the applied field in the $(\theta, \phi)$ direction can be written as
\begin{equation}
\label{eq:Gamma}
\Gamma (t; \, \theta, \phi) = \sum_{ij} Q^{(2)}_{ij}  \left[ \gamma_{2x} (t)  \frac{\partial B_{i} }{\partial x} \frac{\partial B_{j} }{\partial x} + \gamma_{2y} (t) \frac{\partial B_{i} }{\partial y} \frac{\partial B_{j} }{\partial y} \right]
\end{equation}
where $\gamma_{2x} (t)$ and $\gamma_{2y} (t)$ are the prefactors related to the gradients in $x$- and $y$-direction respectively.  They are calculated as follows: for the UD model
\begin{equation}
    \gamma_{2x} (t) = \gamma_{2y} (t) = \; (\frac{2\mu_B}{\hbar})^2 (\frac{q}{2m w_{orb}^2})^2  \frac{\rho_v p_0^2}{576 \pi \epsilon_0^2} \left( \frac{1}{l^3} - \frac{1}{d^3} \right) 2\pi \tau (t+(e^{-t/\tau}-1)\tau),
\label{eq:UDgamma}
\end{equation}
and for the UT model
\begin{equation}
\begin{aligned}
    \gamma_{2x} (t) = \; &(\frac{2\mu_B}{\hbar})^2 (\frac{q}{2m w_{orb}^2})^2 \frac{(9\pi+6) \rho_a p_0^2}{512 \pi^2 \epsilon_0^2 d^4} 2\pi \tau (t+(e^{-t/\tau}-1)\tau), \\
    \gamma_{2y} (t) = \; &(\frac{2\mu_B}{\hbar})^2 (\frac{q}{2m w_{orb}^2})^2 \frac{(9\pi-6) \rho_a p_0^2}{512 \pi^2 \epsilon_0^2 d^4} 2\pi \tau (t+(e^{-t/\tau}-1)\tau).
\end{aligned}
\end{equation}
Here $p_0=2.0\times 10^{-29}\,$C$\cdot$m is the root-mean-square strength of the dipole that is assumed to be the product of elementary charge and the size of an aluminum atom. $\rho_v$ and $\rho_a$ are the volume and areal density of the dipoles and used as fitting parameters for the experimental value $T_2^{(x)} = T_2 (\pi/2, 0) = 840 \,$ns, yielding $v = 2.93 \times 10^{20}\, \mathrm{m}^{-3}$ and $a = 2.66 \times 10^{11}\, \mathrm{m}^{-2}$. Since $T_{\phi} \ll T_1$, we assume $T_2 = T_{\phi}$ by the relation $1/T_2 = 1/2T_1 + 1/T_{\phi}$.

For localized cluster dipole model (CD) and localized cluster trap model (CT), the integration is replaced by explicit summation such that the electric field correlation function is
 \begin{equation}
  \langle E_i(\vec{r},t)  E_j(\vec{r},0) \rangle = 
  \frac{1}{48 \pi^2 \epsilon_0^2} p_0^2 \, g(\omega) 
  \frac{3 R'_i R'_j + \delta_{ij} |\vec{R}'|^2}{|\vec{R}'|^8}
 \end{equation}
 for the CD model and
  \begin{equation}
  \langle E_i(\vec{r},t)  E_j(\vec{r},0) \rangle = \frac{1}{16 \pi^2 \epsilon_0^2}
   p_0^2 \, g(\omega) 
  \frac{9 R_z^{\prime 2} R'_i R'_j - 3R'_z R'_j |\vec{R}'|^2 \delta_{iz} - 3R'_z R'_i |\vec{R}'|^2 \delta_{jz} + |\vec{R}'|^4 \delta_{iz}\delta_{jz} }{|\vec{R}'|^{10}}
 \end{equation}
for the CT model. In these single cluster models, $p_0$ is used as fitting parameter because there are no densities.

\section*{Supplementary Note V: Screening effect of metallic gates}

Let us take into account the screening effect of metallic gates by considering image dipoles. Since an image dipole is formed symmetric to real dipole about the interface between the oxide layer and gates, $R'_z \to 2d - R'_z$. The orientation of the image dipole is reversed in $x$ and $y$ direction such that $p_x \to -p_x$ and $p_y \to -p_y$, which does not have effect on our random dipole and trap model. Assuming there is no correlation between random dipoles/traps and their images, we can get the total correlation function as a sum of each correlation function:
 \begin{equation}
  \langle E_i(\vec{r},t)  E_j(\vec{r},0) \rangle = \langle E_i(\vec{r},t)  E_j(\vec{r},0) \rangle_{re} + \langle E_i(\vec{r},t)  E_j(\vec{r},0) \rangle_{im}
 \end{equation}
where subscripts $re$ and $im$ denote the correlation function from real and image dipoles, respectively.

For the UD model, it would be easier to use Eq. $\ref{dipoleee}$ with two integration regions, $\rho \in [0,\infty),\; \phi \in [0,2\pi),\; z \in [37, 137]\,$nm for $\langle E_i(\vec{r},t)  E_j(\vec{r},0) \rangle_{re} $ and $\rho \in [0,\infty),\; \phi \in [0,2\pi),\; z \in [137, 237]\,$nm for $\langle E_i(\vec{r},t)  E_j(\vec{r},0) \rangle_{im} $. This is equivalent to increasing the region of integration to $\rho \in [0,\infty),\; \phi \in [0,2\pi),\; z \in [37, 237]\,$nm and then using Eq. $\ref{dipoleee}$. To exploit the formula in Eq. \ref{eq:Gamma} and Eq. \ref{eq:UDgamma}, the parameters should be changed to $l=200\,$nm and $d=237\,$nm. For the UT model, $\langle E_i(\vec{r},t)  E_j(\vec{r},0) \rangle_{im} = \langle E_i(\vec{r},t)  E_j(\vec{r},0) \rangle_{re} $ because traps are distributed at the interface. Thus, the correlation function for the UT model is effectively increased by a factor of 2.

For the CD model,
 \begin{equation}
  \langle E_i(\vec{r},t)  E_j(\vec{r},0) \rangle = 
  \frac{1}{48 \pi^2 \epsilon_0^2} p_0^2 \, g(\omega) 
  \left( \frac{3 R'_i R'_j + \delta_{ij} |\vec{R}'|^2}{|\vec{R}'|^8} + \frac{3 R''_i R''_j + \delta_{ij} |\vec{R}''|^2}{|\vec{R}''|^8} \right)
 \end{equation}
with $\vec{R}'' = (R'_x, R'_y, 2d - R'_z)$ and for the CT model,
  \begin{equation}
  \langle E_i(\vec{r},t)  E_j(\vec{r},0) \rangle = 
   \frac{1}{8 \pi^2 \epsilon_0^2} p_0^2 \, g(\omega) 
  \frac{9 R_z^{\prime 2} R'_i R'_j - 3R'_z R'_j |\vec{R}'|^2 \delta_{iz} - 3R'_z R'_i |\vec{R}'|^2 \delta_{jz} + |\vec{R}'|^4 \delta_{iz}\delta_{jz} }{|\vec{R}'|^{10}}.
 \end{equation}

\section*{Supplementary Note VI: Hyperfine noise}

The hyperfine noise contribution for dephasing rate is calculated: $1/T_2^{hyper} = (1.83\upmu$s$)^{-1} - (20.4\upmu$s$)^{-1} = (2.01\,\upmu$s$)^{-1}$. This comes from the difference in decoherence rates of natural silicon device and isotopically purified silicon device \cite{Takeda:2016p1600694,Yoneda:2018p102}. Since it is isotropic, it should be added to all the $T_2 (\theta, \phi)$. Hence, $1/T_2 (\pi/2, 0) + 1/T_2^{hyper}$ is set to $(840\,\mathrm{ns})^{-1}$ by fitting the densities or dipole strength.

The equations in this section are used to generate the anisotropy maps of $T_2$ in the main text.

\end{document}